\documentclass[english,prd,superscriptaddress,nofootinbib,preprintnumbers,twocolumn,showpacs]{revtex4}
\usepackage[utf8]{inputenc}
\usepackage[english]{babel}
\usepackage{amsmath}
\usepackage{amsfonts}
\usepackage{amssymb}
\usepackage{epsfig}
\usepackage{graphics,psfrag,rotating}
\usepackage{graphicx}
\usepackage{dcolumn}
\usepackage{bm}
\usepackage{epstopdf}
\usepackage{color}
\usepackage[usenames,dvipsnames,svgnames]{xcolor}
\usepackage[colorlinks=true,
            linkcolor=red,
            urlcolor=gray,
            citecolor=blue]{hyperref}
  \usepackage{hyperref}

\newcommand{\lsim}   {\mathrel{\mathop{\kern 0pt \rlap
{\raise.2ex\hbox{$<$}}}
 \lower.9ex\hbox{\kern-.190em $\sim$}}}
\newcommand{\gsim}   {\mathrel{\mathop{\kern 0pt \rlap
{\raise.2ex\hbox{$>$}}}
\lower.9ex\hbox{\kern-.190em $\sim$}}}

\def\3nab{\tilde{\nabla}}

\def\hsp5{\hspace{5mm}}
\newcommand{\sfrac}[2]{{\textstyle{#1\over#2}}}
\def\case#1/#2{\textstyle\frac{#1}{#2}}

\def\ber {\begin{eqnarray}}
\def\eer {\end{eqnarray}}
\def\bea {\begin{eqnarray}}
\def\eea {\end{eqnarray}}

\def\bc {\begin{center}}
\def\ec {\end{center}}
\def\case#1/#2{\frac{#1}{#2}}

\newcommand{\bw}{\begin{widetext}}
\newcommand{\ew}{\end{widetext}}

\newcommand{\be}{\begin{equation}}
\newcommand{\bse}{\begin{subequation}}
\newcommand{\ese}{\end{subequation}}
\newcommand{\ee}{\end{equation}}
\newcommand{\eei}{\end{eqnarray}\indent\indent}
\newcommand{\ba}{\begin{array}}
\newcommand{\ea}{\end{array}}
\newcommand{\bal}{\begin{eqnarray}}
\newcommand{\eal}{\end{eqnarray}}

\def\case#1/#2{\textstyle\frac{#1}{#2} }

\newcommand{\bver}{\begin{verbatim}}
\newcommand{\ever}{\end{verbatim}}

\begin{document}
\title{Cosmological dynamics of viable $f(R)$ theories of gravity}
\author{Sulona Kandhai\footnote{kndsul001@myuct.ac.za}}
\affiliation{Astrophysics, Cosmology and Gravity Centre (ACGC), Department of Mathematics and Applied Mathematics, University of Cape Town, Rondebosch 7701, Cape Town, South Africa.}
\author{Peter K. S. Dunsby\footnote{peter.dunsby [at] uct.ac.za}}
\affiliation{Astrophysics, Cosmology and Gravity Centre (ACGC), Department of Mathematics and Applied Mathematics, University of Cape Town, Rondebosch 7701, Cape Town, South Africa.}
\affiliation{South African Astronomical Observatory,  Observatory 7925, Cape Town, South Africa}
\begin{abstract} 
A complete analysis of the dynamics of the Hu-Sawicki modification to General Relativity is presented. In particular, the full phase-space is given for the case in which the model parameters are taken to be $n=1, c_{1}=1$ and several stable de Sitter equilibrium points together with an unstable ``matter-like" point are identified.

We find that if the cosmological parameters are chosen to take on their $\Lambda$CDM values today, this results  in a universe which, until very low redshifts, is dominated by an equation of state parameter equal to $\frac{1}{3}$, leading to an expansion history very different from $\Lambda$CDM. We demonstrate that this problem can be resolved by choosing $\Lambda$CDM initial conditions at high redshift and integrating the equations to the present day. 
\end{abstract}
\maketitle
\section{Introduction}
Over the past decade, our ability to make high precision measurements of the cosmic microwave background and distance measurements of Type Ia supernovae have led to the remarkable conclusion that the expansion rate of the universe is accelerating. 

The most widely accepted cosmological model of the universe, based on General Relativity -- {\em The Concordance Model} -- attributes this acceleration to a mysterious, exotic form of energy density called {\em dark energy}, which dominates the energy density of the universe at the present time. While this model is by far the simplest and most successful, in terms of fitting the current cosmological data, there are a number of shortcomings associated with this theory, such as fine tuning issues, the coincidence problem and the need for unknown, exotic forms of energy.

Naturally, such fundamental problems begs the question of whether or not our understanding of physics is complete on the scales which we are applying it; for example, could the late time acceleration of the universe be a manifestation of the break down of the gravitational interaction on cosmological scales? It is for this reason that viable alternative theories of gravitation are currently receiving a great deal of attention. 

A very popular class of alternatives to the $\Lambda$CDM model involve modifying, the Einstein-Hilbert action by considering terms containing higher order curvature invariants. This results in a gravitational theory which produces predictions that differ from General Relativity \textit{only} late in the matter dominated era of the universe, producing an effective dark energy term which leads to the observed late time acceleration, as well as retaining the success of General Relativty on solar system scales. These $f(R)$ theories of gravity, by their construction, introduce an additional scalar degree of freedom which has a coupling to matter that is extremely weak \cite{Beyond},\cite{Hu2008},\cite{Cardone}. This scalar degree of freedom results in a long range fifth force which predicts that the curvature of the space time in the neighbourhood of a localised energy density is uncoupled from the energy density.  Taking this effect into consideration, several conditions must be satisfied in order for a $f(R)$ theory to be compatible in both the low and high curvature regimes \cite{Hu2008}. It is important that the expansion history in the early universe matches what is obtained from General Relativity, in order to be consistent with big bang nucleosynthesis, as well as the fact that the $f(R)$ theory should generate a late time acceleration compatible with current observations, without the introduction of a cosmological constant. 

If we take the function $f(R)$ to be made up of the usual linear Lagrangian plus some corrective terms encapsulated in a second function $g(R)$: $f(R) = R+g(R)$, then the above conditions can be expressed as the following limits on $g(R)$: $\lim\limits_{R\rightarrow\infty} g(R)= {\textstyle Constant}\;,~
\lim\limits_{R\rightarrow 0} g(R) = 0$. The first limit corresponds to epochs during which the model $f(R)$ should resemble a late time $\Lambda$CDM universe, containing an effective cosmological constant, for which the Lagrangian looks like $R-2\Lambda$. The second limit enforces the condition that the $f(R)$ model chosen must behave close to standard GR. 

So far, the most compelling models proposed, which satisfy these limits are broken power law functions, due to the fact that such functions can be parametrised to give the required behaviour in the relevant regimes. Several models have been proposed which are designed to satisfy the above limits \cite{Hu2008, Starobinsky:2007} and we will focus on the first of these, introduced originally by Hu \& Sawicki \cite{Hu2008}.

The main difficulty which plagues the investigation of such models results from the inherent complexity associated with the inclusion of higher order invariants of the curvature tensor into the gravitational action. It is consequently extremely difficult if not impossible to derive the cosmological dynamics and obtain exact solutions for such theories. 

One method, which has been found to be particularly effective in overcoming many of the issues encountered with higher order theories, is the so-called dynamical systems approach to cosmology \cite{Carloni2007, Carloni2009, shosho2012}. This technique relies on the ability to express the cosmological equations as a set of autonomous differential equations, using a set of suitably chosen variables. In this paper we use these techniques to proved a detailed description of the cosmological dynamics of the Hu \& Sawicki model and determine under what conditions a $\Lambda$CDM-like expansion history can be obtained. 

In Section \ref{sect-f(R)} below we briefly present the field equations for $f(R)$ gravity in a metric formalism. Section \ref{sect-HS} discusses the Hu-Sawicki model, which is the focus of analyses performed below. We present the dynamical systems approach to $f(R)$ gravity, give the general form of the dynamical system for a general $f(R)$ and carefully detail the analysis and its results in Section \ref{sect-DS}, where we discuss the fixed points, and their stability, both in a compact and non-compact phase space. The expansion history for the particular Hu-Sawicki model, considered in this paper  is calculated in Section \ref{sectExpHist} to provide a comparison with the Concordance model, as well as to illustrate the effects of fixing the initial conditions on the viability of the model. Finally, we review the results presented and discuss conclusions and various limitations of the analysis in Section \ref{sect-conc}. Throughout this paper natural units are assumed ($\hbar = c = k_{B}=8\pi G =1$). 

\section{$f(R)$ Cosmology} \label{sect-f(R)}
In principle there is no a-priory reason to restrict the gravitational Lagrangian. In fact, it is quite possible that the addition of higher powers of $R$ and corresponding invariants may improve the characterisation of gravitational fields near regions where $R\rightarrow \infty$ \cite{barrow}. Further impetus to explore results of nonlinear gravitational Lagrangians is due to the fact that every theory attempting to unify the fundamental interactions require either that there are non minimal couplings to the geometry or that higher order curvature invariants appear in the action \cite{Beyond},\cite{barrow}. 

In higher order gravity theories, the Einstein-Hilbert action is modified by considering the addition of higher order, most commonly, \textit{second order}, curvature invariants. Second order modifications result in quadratic Lagrangians, which involve some of the four possible curvature invariants of the Ricci scalar, Ricci and Riemann tensors; $R^{2}, R_{ab}R^{ab},R_{abcd}R^{abcd}$ and $\varepsilon^{iklm}R_{ikst}R^{st}_{lm}$, here $\varepsilon^{iklm}$ is the completely antisymmetric tensor, of rank four. In fact, by making use of the following identities, which are true for all 4-dimensional space times:
\begin{align}
(\delta / \delta g_{ab}) & \int d^{4}x \sqrt{-g}(R^{abcd}R_{abcd}-4R^{ab}R_{ab}+R^{2})=0\;,\\
(\delta / \delta g_{ab}) & \int d^{4}x \sqrt{-g}\varepsilon^{abcd}R_{abcd}R_{cd}^{ef}=0\;,
\end{align}
as well as the fact that the following identity is true for spacetimes with maximally symmetric spatial sections,
\begin{flushleft}
\begin{equation}
(\delta / \delta g_{ab})  \int d^{4}x \sqrt{-g} (3R^{ab}R_{ab}-R^{2})=0\;,
\end{equation}
\end{flushleft}
it is possible to deduce that any quadratic modifications to the Lagrangian can in general be represented by combinations of powers of the Ricci scalar. The modified Lagrangian is then given by:
\begin{equation}\label{modlagrangian}
L = \sqrt{-g} \left( f(R)+\mathcal{L}_{m} \right)\;,
\end{equation}
where $\mathcal{L}_{m}$ is the standard matter Lagrangian. Clearly General Relativity is recovered when $f(R)=R-2\Lambda$. Varying (\ref{modlagrangian}), with respect to the metric we obtain the following fourth order field equations:
\begin{equation}
G_{\alpha \beta} + f'R_{\alpha \beta} - \left( \sfrac{1}{2}f - \square f' \right)g_{\alpha \beta} - \nabla_{\alpha}\nabla _{\beta}f' = T_{\alpha \beta}\;.
\end{equation}
Here primes denote derivatives with respect to the Ricci scalar. We are only concerned with modifications to GR at late times, so $T_{\alpha \beta}$ is described completely by the matter dominated stress-energy tensor.  For the spatially flat ($k=0$) FLRW metric with signature $(-+++)$, where the matter is described as a perfect fluid with equation of state $w=\frac{p}{\rho}$, the independent field equations in terms of the function $f(R)$ are (dots represent temporal derivatives):\\

\noindent \textit{The Raychaudhury equation}
\begin{eqnarray}\label{modraych2}
2\dot{H}+3H^{2}&=&-\sfrac{1}{f'}\left( p_{m} + 2H\dot{f'} + \sfrac{1}{2}\left( f-Rf' \right)\right.\nonumber\\
&+&\left.\dot{R}^{2}f''' + \ddot{R}f'' \right)\;,
\end{eqnarray} 
where $H$ represents the usual expansion rate, $H=\frac{\dot{a}}{a}$,\\

\noindent\textit{the Friedmann equation},

\begin{align}\label{modfried2}
{H^{2} = \sfrac{1}{3f'}\left( \rho_{m} + \sfrac{1}{2}\left( Rf'-f \right)-3H\dot{f'} \right)}.
\end{align}

\noindent \textit{The trace equation}

\begin{align}\label{modtrace2}
3\ddot{R}f'' = \rho(1-3w) + f'R-2f-9Hf''\dot{R}-3f'''\dot{R}^{2},
\end{align}
In addition the {\em Energy Conservation equation} for standard matter is given by
\begin{equation}
\dot{\rho}=-3H(1+w)\rho.
\end{equation}
Finally an expression for the Ricci scalar in terms of $H$ can be obtained by combining the Raychaudhuri and the Friedmann equations:
\begin{equation}
R=6\dot{H}+12H^{2}.
\end{equation}
\section{The Hu-Sawicki Model}\label{sect-HS}
The Hu-Sawicki model (HS model) for $f(R)$-gravity \cite{Hu2008} was designed specifically to overcome several problems faced by other $f(R)$ functions considered as alternatives to the Lagrangian of General Relativity. It is constructed in such a way that it satisfies the limits mentioned in the previous section, and therefore, given an appropriate choice of parameters, is able to recover standard General Relativity at high redshifts, as well as mimic $\Lambda$CDM at low redshifts. See \cite{Hu2008} for details on this class of models. Below is a short summary of the parameters which enter this theory. The functional form of $f(R)$ is characterised by the correction $g(R)$, which is a general class of broken power laws:
\begin{equation}\label{hsmodel}
f(R) = R - m^{2}\frac{c_{1}\left(\frac{R}{m^{2}}\right)^{n}}{c_{2}\left(\frac{R}{m^{2}}\right)^{n}+1}\;.
\end{equation}
Here $n>0$ and the mass scale $m^{2}$ is taken to be:
\begin{equation}\label{msq}
m^{2}\equiv \frac{\kappa^{2}\bar{\rho}_{0}}{3}=(8315Mpc)^{-2}\left( \frac{\Omega_{m}h^{2}}{0.13} \right)\;,
\end{equation}
where $\bar{\rho}_{0}$ is the average density of the present epoch and $\kappa^{2}=8\pi G$. $c_{1}$ and $c_{2}$ are dimensionless parameters, whose relationship will be shown to be associated with the present matter density parameter, $\Omega_{m}$.

The sign of the second derivative of the correction, $g(R)$,  is required to be strictly positive for high curvatures relative to density in order to guarantee that at high density, the solution is stable in this regime. 
\begin{equation}
g_{RR} \equiv \frac{d^{2}g(R)}{dR^{2}} >0 \text{~~~for~~~}R \gg m^{2}\;.
\end{equation}
The above condition is important to ensure that at high redshifts, the General Relativity values of quantities, such as $\Omega_{m}h^{2}$ are recovered.  For convenience we define $g_{R} \equiv dg/dR$.

The key appeal of this class of models is that, explicitly, there is no cosmological constant term. However, if we consider the limit of very high curvature relative to $m^{2}$: 
\begin{equation}\label{lim}
\lim\limits_{m^{2}/R\rightarrow 0} f(R) \approx R-\frac{c_{1}}{c_{2}}m^{2}+\frac{c_{1}}{c_{2}^{2}}m^{2}\left( \frac{m^{2}}{R} \right)^{n}\;,
\end{equation}
it can be seen that, for a fixed value of $\frac{c_{1}}{c_{2}}$, the limiting case of $\frac{c_{1}}{c_{2}^{2}}\rightarrow 0$ behaves like a cosmological constant for both local and cosmological scales. At finite values of $\frac{c_{1}}{c_{2}^{2}}$, the curvature is frozen to a fixed value and no longer decreases with matter density; this gives rise to a set of models which exhibit late time acceleration similar to $\Lambda$CDM \cite{Hu2008}.

It follows that \cite{Hu2008} admits the following relationship between the parameters and the energy densities of the cosmological constant, $\tilde{\Omega}_{\Lambda}$, and matter, $\tilde{\Omega}_{m}$:
\begin{equation}\label{ratio c1c2}
\frac{c_{1}}{c_{2}} \approx 6\frac{\tilde{\Omega}_{\Lambda}}{\tilde{\Omega}_{m}}\;.
\end{equation}
The remaining two parameters $n$ and $\frac{c_{1}}{c_{2}^{2}}$  are free to choose in order to determine the expansion history and how well it fits observations, compared to the $\Lambda$CDM model. Larger values of $n$ allows the model to mimic $\Lambda$CDM until later in the expansion history, while smaller values of $c_{1}/c_{2}^{2}$ leads to a reduction in general deviations from it. Because both the Hubble parameter and the critical density depend on the correction $g_{R}$, $\tilde{\Omega}_{m}$ only becomes the actual value of the matter density in the following limit:
\begin{equation}
\lim \limits_{c_{1}/c_{2}^{2}\rightarrow 0} \tilde{\Omega}_{m}=\Omega_{m}\;.
\end{equation}
The matter density in the physical units, however, remains unchanged. 
\subsubsection{Fixing the parameter values today}
Considering the $\Lambda$CDM model with flat spatial geometry, we have:
\begin{equation}
R \approx 3m^{2}\left( \frac{1}{a^{3}}+4\frac{\tilde{\Omega}_{\Lambda}}{\tilde{\Omega}_{m}} \right)\;,
\end{equation}
\noindent and for the derivative of the correction:
\begin{equation}
g_{R} = -n \frac{c_{1}}{c_{2}^{2}}\left( \frac{m^{2}}{R} \right)^{n+1}\;.
\end{equation}
The values for today are:
\begin{equation}\label{initial R and g}
R_{0} \approx m^{2}\left( \frac{12}{\tilde{\Omega}_{m}} - 9 \right)\;,~
g_{R0}  \approx -n\frac{c_{1}}{c_{2}^{2}}\left( \frac{12}{\tilde{\Omega}_{m}} - 9 \right)^{-n-1}
\end{equation}
for $|g_{R0}|\ll1$. \cite{Hu2008} uses the following specific values:
\begin{equation}
\tilde{\Omega}_{m} =0.24\;,~\tilde{\Omega}_{\Lambda}=0.76\;.
\end{equation}
Therefore we have:
\begin{equation}
R_{0}=41m^{2}\;,~g_{R0} \approx -n\frac{c_{1}}{c_{2}^{2}}\left(\frac{1}{41}\right)^{n+1}\;.
\end{equation}
Figure 9 in \cite{Hu2008} show the range and combinations of parameter values,  $g_{R0}$ and $n$, which are acceptable and enable the model to satisfy solar system tests. 
\section{Dynamical Systems approach to $f(R)$ gravity}\label{sect-DS}
We follow the general strategy for applications of the dynamical systems  approach to the cosmology of fourth order gravity, developed in \cite{Carloni2007, Carloni2009, shosho2012, shosho2007}. This provides a way of analysing the dynamics of any analytic function $f(R)$ which is invertible for the Ricci scalar. Following the approach of \cite{shosho2012, Goheer2007, Goheer2008}, a compact analysis of the phase space for a general $f(R)$ theory is performed, by defining a strictly positive normalisation to pull any solutions at infinity into a finite volume.
\begin{figure}[htb!]
{\label{phase2}\includegraphics[width=0.4\textwidth,angle=0]{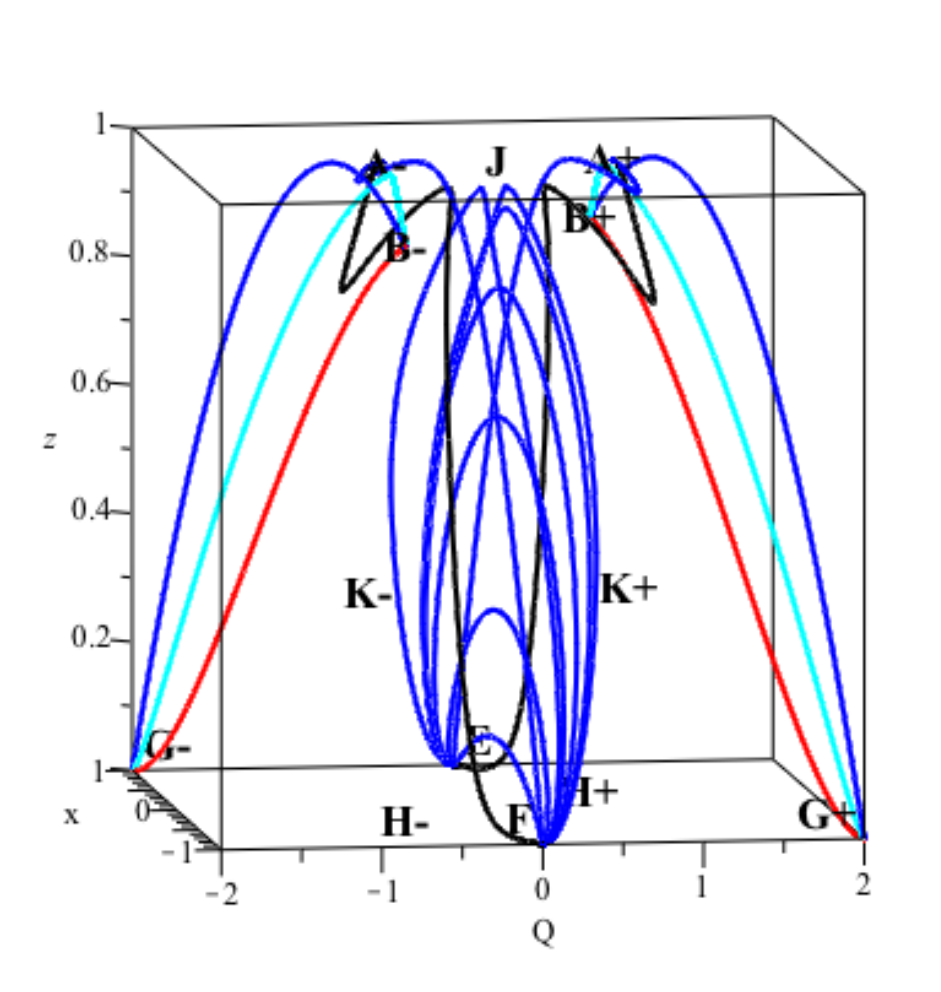}}
{\label{v0invsub}\includegraphics[width=0.4\textwidth,angle=0]{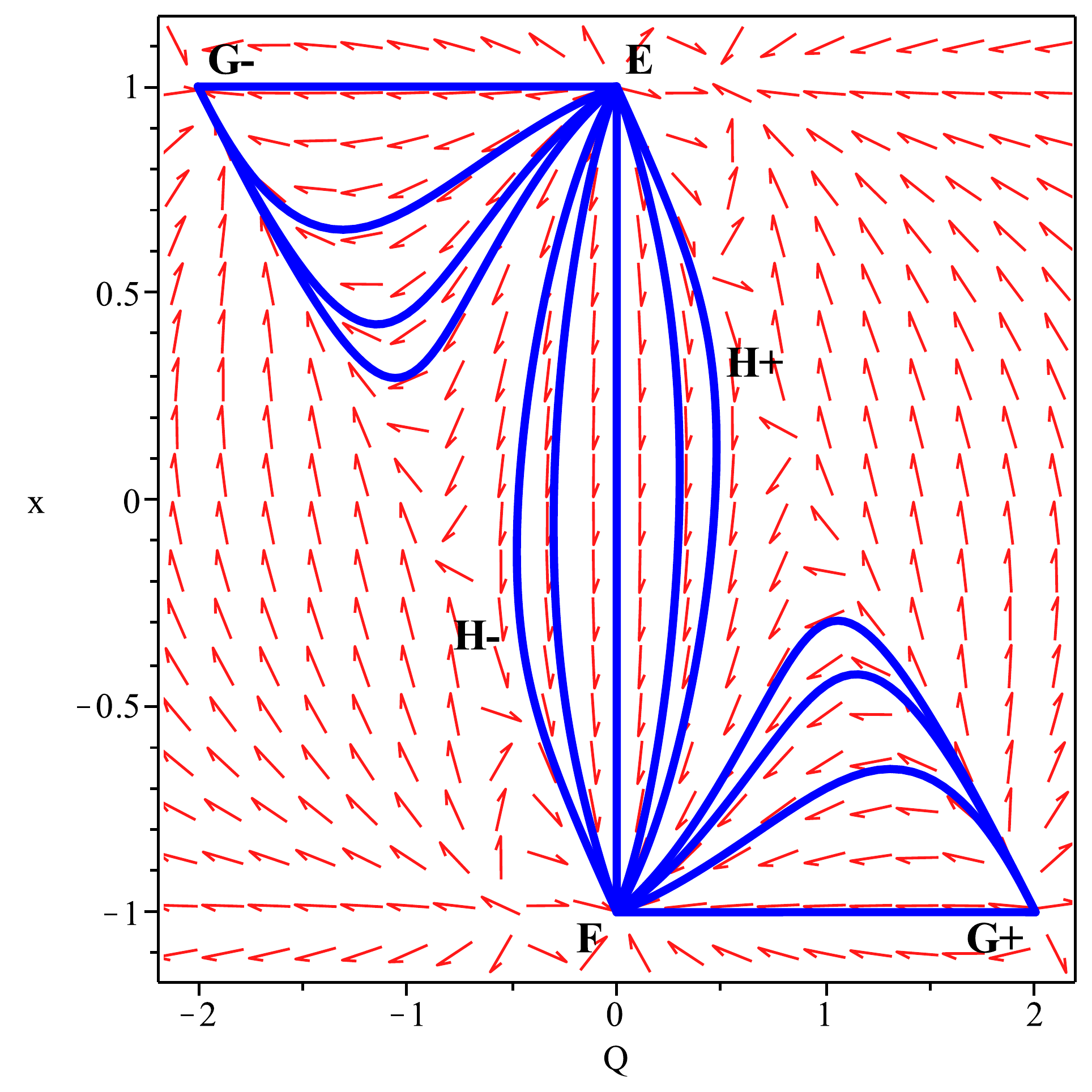}}
\caption[justification=justified,singlelinecheck=false]{{\bf Top panel} - this gives a perspective of the 3-dimensional compact phase space of the Hu-Sawicki model for $n=1, c_{1}=1$. This plot demonstrates the anti-symmetry between the expanding and contracting regions of the phase space, defined by the $Q=0$ plane, as well as the possibility that orbits can cross this plane --  implying the existence of universe models with bounce behaviour. {\bf Bottom panel} - This gives the phase plane of the invariant submanifold $v=0$, corresponding to universe models with zero Ricci curvature.}
\label{Figure 5}
\end{figure}
In order to make this possible, an appropriately normalised time variable needs to be defined. Ensuring that this time variable is strictly non-decreasing amounts to guaranteeing that the expansion-normalisation adopted is strictly positive. Thus, the sign of time is maintained, whether studying expanding or collapsing cosmologies.  This means that any negative contribution to the Friedmann equation must be absorbed into the normalisation. In the case where a quantity can be both positive and negative, each option must be studied in a separate sector of the phase space. The full phase space can then be reconstructed by simply aligning the various sectors along their common boundaries. A phase space which is constructed in this way can therefore include all static, bouncing and re-collapsing models.
 
In this paper we will only consider sectors of the phase space where $R \geq 0$, simply because negative Ricci scalar values are not of any real physical interest. We require $f',f''>0$, and the matter density is assumed to be non-negative. The function $f$ under consideration here, is always positive. Therefore, in (\ref{modfried2}), the term which contains $f$ attached to a negative sign must be absorbed into the positive-definite normalisation. 
\subsection{Compact phase space analysis of $f(R)$ gravity} \label{compAnalysis}
Rewriting the modified Friedmann equation (\ref{modfried2}) in the following way allows a convenient definition of the normalised variables:
\begin{align}\label{friedrewrite}
\left( 3H + \frac{3}{2} \frac{\dot{f'}}{f'} \right)^{2}+\frac{3}{2}\frac{f}{f'} = \frac{3\rho_{m}}{f'} &+ \frac{3}{2}R +\left( \frac{3}{2}\frac{\dot{f}}{f'} \right)^{2}\;.
\end{align}
The left hand side of the above equation is a positive definite quantity. Quite naturally, assigning each term in (\ref{friedrewrite}) a name, we obtain the following dynamical variables:
\begin{align}\label{variables}
x&=\frac{3}{2} \frac{\dot{f'}}{f'}\frac{1}{D},~~~~v=\frac{3}{2}\frac{R}{D^{2}},\nonumber \\
&\\ 
y&=\frac{3}{2}\frac{f}{f'}\frac{1}{D^{2}}, ~~~~\Omega =\frac{3\rho_{m}}{f'}\frac{1}{D^{2}}\;,
~~~~Q=\frac{3H}{D},       \nonumber                       
\end{align}
where $D$ represents the normalisation which compactifies the phase space, and takes the form 
\begin{equation}\label{defD}
D^{2}=\left(3H + \frac{3}{2}\frac{\dot{f'}}{f'}\right)^{2} + \frac{3}{2}\frac{f}{f'}\;.
\end{equation}
The normalised time variable is then defined as follows:
 \begin{equation}\label{ntime}
\frac{d}{d\tau} \equiv \frac{1}{D} \frac{d}{dt}~.
\end{equation}
Thus, (\ref{friedrewrite}) and (\ref{defD}) establish two independent constraint equations for our system:
\begin{align}
1&=\Omega + x^{2} +v\;, \label{friedcon}\\
&\nonumber\\
1& = (Q+x)^{2} + y\;.\label{normcon}
\end{align}
The dynamical variables defined by (\ref{variables}) constitute the coordinates of our compact phase space and the boundaries of this phase space are defined by the above two constraints as follows:
\begin{align}
-1 \leq x\leq 1, ~~~&~ 0 \leq \Omega \leq 1, ~~~~ -2 \leq Q \leq 2\;,\nonumber\\
\\
0 \leq&  v \leq 1,~~~~ 0 \leq y \leq 1\;. \nonumber
\end{align}
The construction of $D$ ensures that the dynamical variables are well defined when $H=0$, thus we expect all static, expanding, collapsing and bounce solutions to be included. Expanding and collapsing universes are connected across the $Q=0$ boundary.
\subsection{The General Propagation Equations}
Differentiating the dynamical variables (\ref{variables}) with respect to  $\tau$ and substituting the independent cosmological equations, (\ref{modfried2})$-$(\ref{modtrace2}), leads to a set of 5 first order autonomous differential equations. The dimensionality of the system can be reduced by using the constraint equations (\ref{friedcon}) and (\ref{normcon}) to eliminate $y$ and $\Omega$. Below, we show the general propagation equations for the  reduced 3-dimensional autonomous system, where 
\begin{equation}
\Gamma \equiv \frac{f'}{f''R}
\end{equation}
specifies the specific $f(R)$ model. In order to close the system, $\Gamma$ must be expressed in terms of the dynamical variables. This implies that the above system characterises a general dynamical system for any modified gravity cosmology defined by a function $f(R)$, which is invertible in terms of the dynamical variables, in other words,  $\Gamma$ needs to be expressed as a function of $(x,Q,v)$:
\begin{widetext}
\begin{align}
\frac{dv}{d\tau} =& -\frac{1}{3}v \left[  \left( Q+x \right)  \left( 2\,v+4\,xQ- \left( 1-v-{x}^
{2} \right)  \left( 1+3\,w \right)  \right) \label{eqv}
  -2\,Q-4\,x+2\,x{\Gamma}\,\left( v-1 \right)  \right]\;,\\ \nonumber
& \nonumber \\
\frac{dx}{d\tau} =& \frac{1}{6}\left[-2\,{x}^{2}v{\Gamma}+\left( 1-v-{x}^{2} \right)  \left( 1-3\,w \right) \label{eqx}
+2v+4\, \left( {x}^{2}-1 \right)  \left( 1-{Q}^{2}-xQ \right) \right. \nonumber\\ 
&\left. +x \left( Q+x \right) \left(  \left( 1-v-{x}^{2} \right)  \left( 1+3\,w \right) -2\,v\right) \right]\;, \\
& \nonumber \\
\frac{dQ}{d\tau} =& \frac{1}{6} \left[ -4\,x{Q}^{3}+ \left( 5+3\,w \right) Qx \left( 1-xQ \right)
 -{Q}^{2} \left( 1-3\,w \right) -Q{x}^{3} \left( 1+3\,w \right) \right. \nonumber \\
& \left. -3\,vQ \left( 1
+w \right)  \left( Q+x \right) +2\,v \left( 1-{\Gamma}\,Qx \right) 
  \right]\;.\label{eqQ}
\end{align}
\end{widetext}
Clearly, $v=0$ corresponds to an invariant submanifold; solutions which start their evolution in this plane will remain there forever. This submanifold corresponds to universe models for which the Ricci scalar vanishes. Due to the existence of an invariant submanifold, no global attractor can exist for cosmological systems defined by the above compact dynamical variables. 

It is useful to express the cosmological equations terms of the dynamical variables. By first expressing the second time derivative of $f'$, given by the trace equation, in terms of these variables:
\begin{equation}\label{vartrace}
\frac{\ddot{f'}}{f'} =\frac{H^{2}}{Q^{2}}\left[ (1-3w)\Omega + 2v -4y -2xQ \right]\;.
\end{equation}
and then using (\ref{vartrace}) and the constraint equations, we can write the modified Raychaudhuri equation (\ref{modraych2}) in the following useful form:
\begin{equation}\label{varraych}
\dot{H}=-\frac{H^{2}}{Q^{2}} \left( 1+\Omega-2\,y-{x}^{2} \right)\;. 
\end{equation}
In this way, once the fixed points have been determined, the expansion history at these points can be found directly by integrating this equation.
\subsection{Dynamical Systems analysis of the Hu-Sawicki model}
Let us now focus on the HS model as described in \cite{Hu2008}, where $f(R)$ is given by:
\begin{equation}\label{hsmodel2}
f(R) = R - m^{2}\frac{c_{1}\left(\frac{R}{m^{2}}\right)^{n}}{c_{2}\left(\frac{R}{m^{2}}\right)^{n}+1}\;.
\end{equation}
We define the following relations:
\begin{equation}\label{dimensionlessR}
m^{2} \equiv \lambda H_{0}^{2}\;,~ r  \equiv \frac{R}{H_{0}^{2}}\;,~ h  \equiv \frac{H}{H_{0}}\;,
\end{equation}
which enables us to write (\ref{hsmodel2}) in a more convenient form:
\begin{equation}\label{hsmodel3}
f(R)=CH_{0}^{2}\left[ \left(\frac{r}{C}\right)-\frac{c_{1}\left(\frac{r}{C}\right)^{n}}{c_{2}\left(\frac{r}{C}\right)^{n}+1} \right]\;.
\end{equation}
Here $c_{1}, c_{2}$ and $n$ are constant parameters to be constrained by observations and the dimensionless parameter $\lambda$ is related to the ratio of $m^{2}$ defined by (\ref{msq}) and the present day value of the Hubble parameter. 

We perform the compact dynamical systems analysis outlined in the previous section is to (\ref{hsmodel2}) or (\ref{hsmodel3}) for the case $n=1$. It has been shown that $n$ remains unconstrained by current cosmological data \cite{Santos2012} and, therefore does not compromise generality at this stage. For simplicity we also set $c_{1}=1$. This has the added benefit of leading to a $\Gamma$ which can be expressed entirely in terms of the dynamical systems variables:
\begin{equation}\label{gamma}
\Gamma \equiv \frac{f'}{f''R} = \frac{1}{2}\,{\frac {vy}{ \left( v-y \right) ^{2}}}~.
\end{equation} 
The expression for $\Gamma$ associated with this specific model (\ref{gamma}) is then substituted into equations (\ref{eqv}) - (\ref{eqQ}), and a dynamical systems analysis is performed to obtain the equilibrium points of the system. The equilibrium points, as well as their corresponding exact solutions for the scale factor are given in Table \ref{sols}. The Hartman-Grobman theorem is used to determine the stability of these fixed points, where possible. Some points obtained are non-hyperbolic, and in this case we resort to the Center Manifold Theorem or other techniques to clarify their nature. 
In Figure \ref{Figure 5}, we present a view of the compact phase space by showing several trajectories between the fixed points; in the upper panel we show the entire 3D phase space, and in the lower panel we show the invariant sub manifold corresponding to $v=0$.
\subsection{Stationary points, stability and exact solutions}
For the HS model, with $n=c_{1}=1$ and keeping an arbitrary equation of state  $w$, the fixed points for the entire phase space, as well as the exact solution of the scale factor at each point, are summarised in Table \ref{sols}.

Note that although Table \ref{table:stability} includes a stability classification of the fixed points identified for universes dominated respectively by radiation, matter and a cosmological constant, the bulk of the analysis is focused on universes dominated by a dust fluid ($w=0$). Accordingly, the phase-portraits below are all generated for $w=0$.

The points for which the $v$ and/or $y$ coordinates are exactly zero make up a subset of fixed points which appear in the dynamical systems analysis for any arbitrary $f(R)$ theory of gravity since $\Gamma=0$ and the resulting equations are independent of $f(R)$. For example, we find the corresponding fixed points for which $v=0$ that appear in \cite{shosho2012} for $f(R)=R+\alpha R^{n}$.
\begin{table*}
\centering
\begin{tabular}{l l c c} \hline \hline
Point &  Coordinate $(Q,x,y,v,\Omega)$& Scale factor evolution, $a(t)$ \\ \hline
& & & \\
$\mathcal{A}_{\pm}$  & $\left[\pm\frac{1}{\sqrt{2}},0,\frac{1}{2},1,0\right]$& $a(t)=a_{0}e^{H_{0}(t-t_{0})}$  \\
$\mathcal{B}_{\pm}$  & $\left[\pm\frac{2}{3},\pm\frac{1}{3},0,\frac{8}{9},0 \right]$&$a(t)=a_{0}e^{H_{0}(t-t_{0})}$\\
$\mathcal{C} (w=-1)$   & $\left[ 0,\sqrt{\frac{3(w+1)}{1+3w}},-\frac{2}{1+3w},0,-\frac{2}{1+3w} \right]$&$a(t)=a_{0}$\\
$\mathcal{D} (w=-1)$   & $\left[ 0,-\sqrt{\frac{3(w+1)}{1+3w}},-\frac{2}{1+3w},0,-\frac{2}{1+3w} \right]$&$a(t)=a_{0}$\\
$\mathcal{E}$  & $\left[ 0,1,0,0,0 \right]$ & $a(t)=a_{0}$\\
$\mathcal{F}$  & $\left[0,-1,0,0,0\right]$  & $a(t)=a_{0}$\\
$\mathcal{G}_{\pm}$  & $\left[ \pm2,\mp1,0,0,0 \right]$&  $ a(t)=a_{0}\left( 2H_{0}(t-t_{0})+1 \right)^{\frac{1}{2}} $\\
$\mathcal{H}_{\pm}(w \leq \frac{2}{3})$ & $\left[  \mp\frac{2}{3(w-1)},\pm\frac{3w-1}{3(w-1)},0,0,-\frac{4}{9}\frac{3w-2}{(w-1)^{2}}  \right]$ &$ a(t)=a_{0}\left( 2H_{0}(t-t_{0})+1 \right)^{\frac{1}{2}} $ \\
$\mathcal{J}$  & $\left[0,0,1,1,0\right]$ & $a(t)=a_{0}$\\
$\mathcal{K}_{\pm}$ &  $\left[\pm\frac{\sqrt{6}}{3},0,\frac{1}{3},\frac{1}{3},\frac{2}{3}\right]$&$a(t)=a_{0}\left(\frac{3}{2}H_{0}\left(t-t_{0}\right)+1\right)^{\frac{2}{3}}$ \\\hline
\end{tabular}
\caption{Each equilibrium point in the phase space has an expanding and collapsing version, denoted by a subscript +/-. Fixed points with $Q=0$, correspond to static universes. The points $\mathcal{H}_{\pm},\mathcal{C}$ and $\mathcal{D}$ depend on the equation of state parameter, $w$; $\mathcal{H}_{\pm}$ only lies in the phase space for $-1 \leq w \leq \frac{2}{3}$, and while $\mathcal{C}$ and $\mathcal{D}$ lie in the phase space for all values of $w \leq -1$. For our purposes $ -1\leq w \leq 1$, therefore we only consider the case when $w=-1$ for these points. The points $\mathcal{C},\mathcal{D},\mathcal{E},\mathcal{F},\mathcal{G}_{\pm},\mathcal{H}_{\pm}$ all lie on the invariant submanifold $v=0$. Other $w$ dependent fixed points were found, but are not included in this analysis as they only lie within the phase space for unphysical values of the equation of state parameter.}
\label{sols}
\end{table*}

\begin{table*}[t]
\centering
\begin{tabular}{l c p{2 cm}| p{2 cm}| p{2 cm}} \hline \hline 
Point               &  Eigenvalues of Jacobian & \multicolumn{3}{c}{Stability} \\ \hline
                    &                          &      $w=0$      & $w=\frac{1}{3}$ &$w=-1$             \\\hline
$\mathcal{A}_{+}$  & $\left[ 0,-\frac{1}{\sqrt{2}},-\frac{1}{\sqrt{2}}-\frac{1}{\sqrt{2}}w \right]$ &\small{Attractor}&\small{Attractor}&\small{Attractor}\\
$\mathcal{A}_{-}$  & $\left[ 0,\frac{1}{\sqrt{2}},\frac{1}{\sqrt{2}}+\frac{1}{\sqrt{2}}w \right]$ &\small{Repellor} &\small{Repellor}&\small{Repellor}\\
$\mathcal{B}_{+}$  & $\left[ -\frac{8}{9},-\frac{1}{9},-\frac{8}{9}-\frac{2}{3}w \right]$ &\small{Attractor}&\small{Attractor}&\small{Attractor}\\ 
$\mathcal{B}_{-}$  & $\left[ \frac{8}{9},\frac{1}{9},\frac{8}{9}+\frac{2}{3}w \right]$ &\small{Repellor}    &\small{Repellor}&\small{Repellor}\\ 
$\mathcal{C}$   & $\left[ \frac{2}{3}\sqrt{\frac{3(1+w)}{1+3w}},-\frac{2}{3}\sqrt{\frac{3(1+w)}{1+3w}},\frac{1}{3}\sqrt{\frac{3(1+w)}{1+3w}} \right]$ & $\square$&$\square$ & \small{Saddle}\\
$\mathcal{D}$   & $\left[ \frac{2}{3}\sqrt{\frac{3(1+w)}{1+3w}},-\frac{2}{3}\sqrt{\frac{3(1+w)}{1+3w}},-\frac{1}{3}\sqrt{\frac{3(1+w)}{1+3w}} \right]$ &$\square$&$\square$& \small{Saddle}\\
$\mathcal{E}$   & $\left[1,\frac{4}{3},\frac{2}{3}\right]$&\small{Repellor}   &\small{Repellor} &\small{Repellor} \\ 
$\mathcal{F}$   &$\left[-\frac{2}{3},-\frac{4}{3},-1\right]$ &\small{Attractor}& \small{Attractor}& \small{Attractor}\\ 
$\mathcal{G}_{+}$  &$\left[\frac{8}{3},3,\frac{4}{3}-2w\right]$ &\small{Repellor}   &\small{Repellor} &\small{Repellor} \\ 
$\mathcal{G}_{-}$  &$\left[-\frac{8}{3},-3,-\frac{4}{3}+2w\right]$ &\small{Attractor}& \small{Attractor}& \small{Attractor}\\ 
$\mathcal{H}_{+}$  &$\left[ -\frac{8}{9(w-1)},-\frac{3w+7}{9(w-1)},-\frac{6w-4}{9(w-1)} \right]$&\small{Saddle}& \small{Saddle}& \small{Saddle}\\ 
$\mathcal{H}_{-}$  &$\left[ \frac{3w+7}{9(w-1)},\frac{8}{9(w-1)},\frac{6w-4}{9(w-1)} \right]$&\small{Saddle}& \small{Saddle}& \small{Saddle}\\ 
$\mathcal{J}$ & $ $& \small{Saddle}& \small{Saddle}& \small{Saddle}\\ 
$\mathcal{K}_{+}$ & $ $ &\small{Unstable spiral} &\small{Unstable} &\small{Unstable} \\
$\mathcal{K}_{-}$ & $ $ &\small{Stable spiral} &\small{Unstable} &\small{Unstable} \\ \hline
\end{tabular}
\caption{In the table above we summarise the stability of each equilibrium point, for equation of state parameters corresponding to dust, radiation and a cosmological constant. $\mathcal{A}_{+}$ and $\mathcal{A}_{-}$ are non-hyperbolic and consequently we use the Center Manifold Theorem to obtain their stability. Points $\mathcal{C}$, $\mathcal{D}$ only exist physically in the phase space for $w=-1$, and the eigenvalues for these points are all equal to zero. The stability of these points were therefore determined by inspection -- they are self-evidently saddle points. The points $\mathcal{J}$ and $\mathcal{K}_{\pm}$ lie on the plane $y=v$; for these points $\Gamma$, and thus the system, is undefined and therefore there are no analytic eigenvalues for these points. The classification of these points was therefore done by inspection. For the cases $w=\frac{1}{3}, -1$, it was only possible to determine whether or not the points were stable.}
\label{table:stability}
\end{table*}
\subsubsection{Stability}
The stability of all the fixed points, save $\mathcal{A}_{\pm}$, $\mathcal{K}_{\pm}$ and $\mathcal{J}$, could be inferred using the Hartman-Grobman theorem. The fixed points $\mathcal{A}_{\pm}$ are non-hyperbolic, so the Center Manifold Theorem was used to determine the stability of $\mathcal{A}_{+}$\footnote{The system is significantly less complicated using the finite variables defined in Section \ref{noncompAnalysis} given by (\ref{ncvariables}), therefore the flow of the Center Manifold at $\mathcal{A}_{+}$ is analysed using this coordinate system.}  Perturbation theory was employed to find the stability near $\mathcal{A}_{-}$. 
\subsubsection{Non analytic points $\mathcal{K}_{+}$ and $\mathcal{J}$}
Considering the structure of $\Gamma$, as expressed in (\ref{gamma}) in terms of the variables, it is clear that when $v=y$, this term becomes undefined. Thus, analytically, it is not possible to identify and analyse stationary points in the phase-space when this occurs. The matter-like points $\mathcal{K}_{\pm}$, which lie on the intersection of the $y=v$ and $x=0$ planes as detailed in \cite{shosho2012} are also present here. The point $\mathcal{J}$ which lies on the $v=y=1$ surface corresponds to a saddle point and represents a static universe. 

The respective stabilities of these points were inferred by inspection; by examining the time evolution of the coordinates, when initial conditions are chosen near the fixed points.

\subsection{Exact solutions}
The rate of expansion, $H$, and the deceleration parameter, $q$,  are related by the following expression:
\begin{equation}\label{hdot}
 \dot{H}=-(1+q)H^{2}\;,  
\end{equation}
where $q = -\frac{\ddot{a}a}{\dot{a}^{2}}$. This relationship can be used to determine the time evolution of the scale factor at each of the equilibrium points, provided the deceleration parameter at that equilibrium point is known.  In order to find the value of $q$ at the $i^{th}$ equilibrium point, we need to express $q$ in terms of the compact variables $x,y,v,Q,\Omega$ . 

It follows, from the Raychaudhuri equation (\ref{varraych}) and the constraint equations, that
\begin{align}
q_{i}=1-\frac{z_{i}}{Q_{i}^{2}}~.
\end{align}
For the cases in which $Q\neq 0$, direct integration of (\ref{hdot}) results in an expression describing the evolution of the scale factor for each equilibrium point of the compact dynamical system. This can be done for the points $\mathcal{A}_{\pm},\mathcal{B}_{\pm},\mathcal{G}_{\pm},\mathcal{H}_{\pm}$, and $\mathcal{K}_{\pm}$. 

Fixed points with deceleration parameter $q=-1$ represent de Sitter universes, with the scale factor evolving exponentially with time, while fixed points for which $q=1$ represent universes which appear to be ``radiation-like"\footnote{Note that this ``radiation-like" behaviour refers to the properties of the curvature fluid -  at these points the curvature fluid causes the expansion of the universe to scale with time in the same way that ordinary radiation would: $a(t) \propto \sqrt{t}$, exhibiting ``radiation-like" behaviour. }, as the scale factor is proportional to the square root of cosmic time. 

One of the short-comings of the dynamical systems approach to cosmology, as outlined in \cite{Carloni2013}, is that it is possible for the analysis to admit fixed points which correspond to solutions of the dynamical system (as defined, for example, by (\ref{eqv}) - (\ref{eqQ})) but \textit{do not} satisfy the cosmological equations. In many cases, constants of integration, which emerge in families of solutions to the cosmological equations, result in additional constraints, which must be satisfied by all physical points of the system. Setting the derivatives of the dynamical variables equal to zero;
\begin{equation}\label{eqx1a}
x' = F(x) = 0
\end{equation}
implies either
\begin{equation}\label{eqx2a}
F(x)=0
\end{equation}
or 
\begin{equation}\label{eqx3a}
x'=0 \Rightarrow x=constant\;.
\end{equation}
Solutions to (\ref{eqx1a}) may result from solving either of the equations (\ref{eqx2a}) or (\ref{eqx3a}), where the latter now represents a set of constraints imposed on the system \cite{Carloni2013}.  For this reason, it is important to verify that the solutions obtained satisfy the cosmological equations.

The points $\mathcal{C},\mathcal{D}, \mathcal{E}$, $\mathcal{F}, \mathcal{J}, \mathcal{K}$, all lie on the non-invariant $Q=0$ submanifold, describing solutions for which the scale factor has no time dependence, and so represent static universes.

For the HS model with $n=c_{1}=1$, we find that the non-static, analytic fixed points belong to one of two scale factor solutions: radiation-like or de Sitter like expansion. The fact that other cosmological evolutions do not appear as stationary phase states does not imply that this model does not allow them -- it may be that the choice of variables places analytic limitations on what can appear as stationary solutions. 

\begin{figure}[htb!]
 {\label{trajB}\includegraphics[width=0.36\textwidth,angle=0]{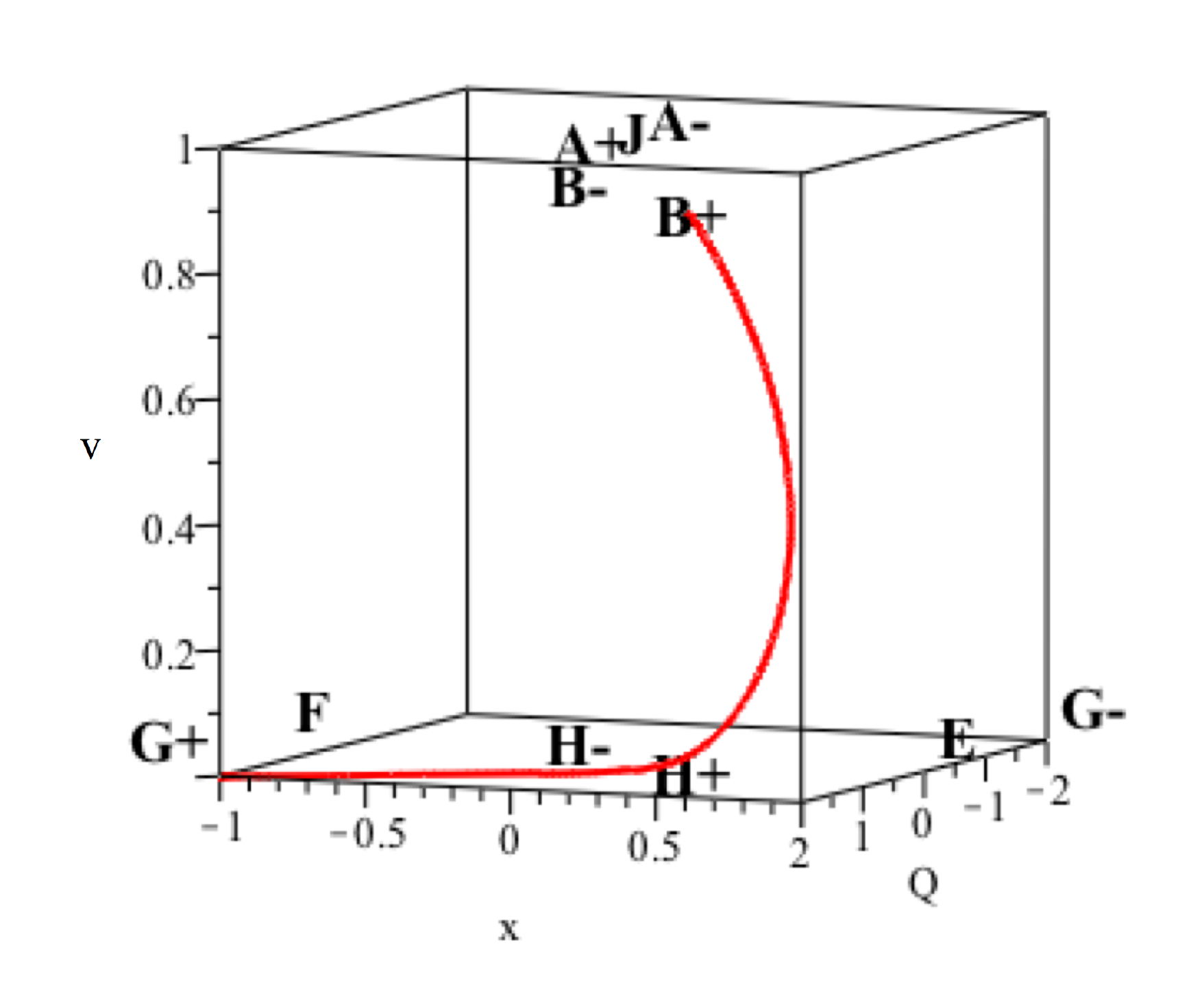}}
{\label{trajA}\includegraphics[width=0.35\textwidth,angle=0]{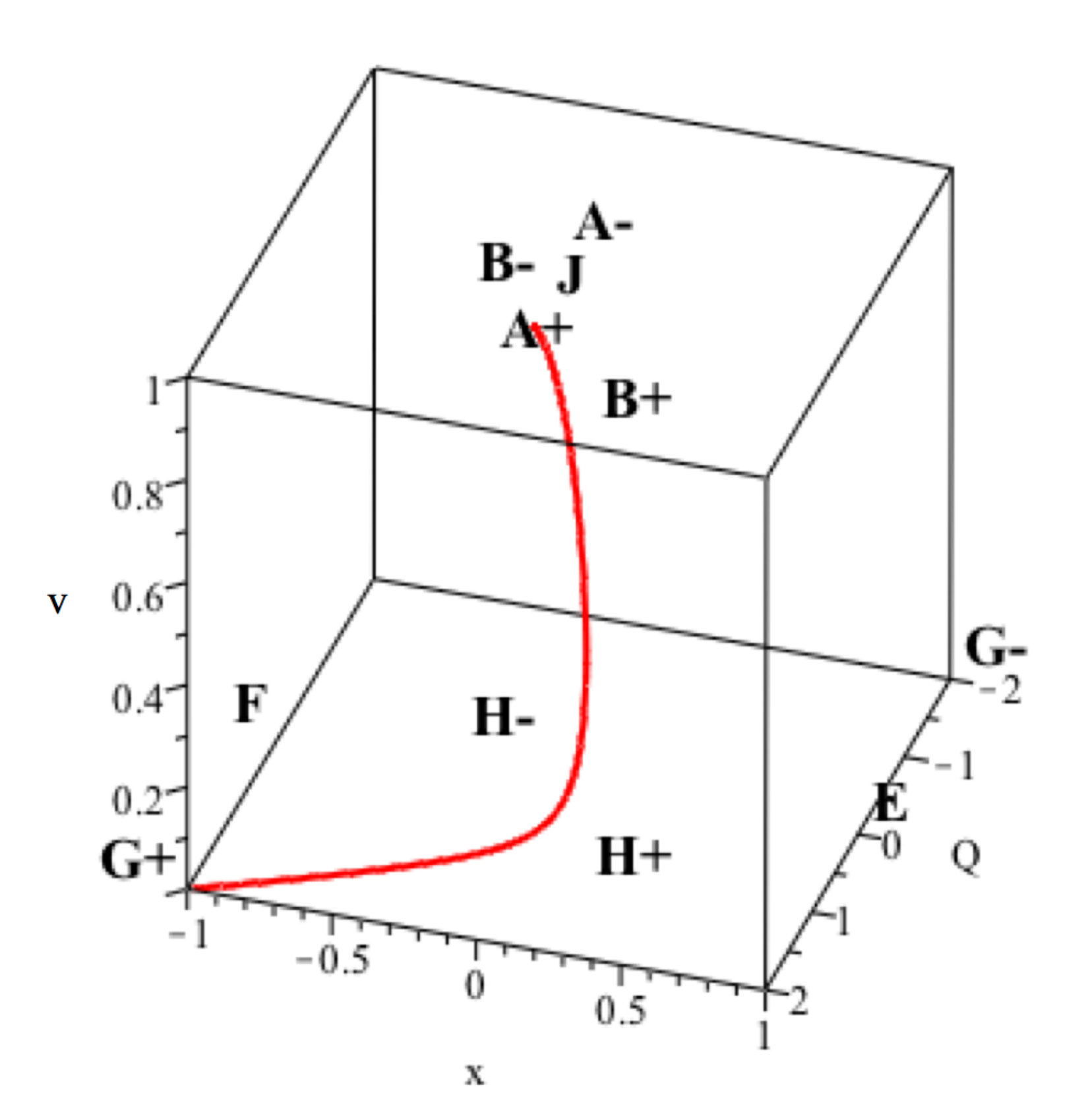}}
\caption{Here two examples of solution orbits are presented. The orbit represented in the left panel begins near the radiation like repeller point $\mathcal{G}_{+}$, is pulled toward the saddle radiation like point $\mathcal{H}_{+}$ and eventually evolves toward the de Sitter like attractor $\mathcal{B}_{+}$.  The orbit represented in the right panel also begins near $\mathcal{G}_{+}$, but evolves towards the de Sitter point, $\mathcal{A}_{+}$. The existence of these orbits show that this model naturally admits solutions which have a late time de Sitter expansion, which can produce expansion histories which look like a Dark Energy fluid is dominating the universe at late times.}
\label{figure:compact orbits}
\end{figure}

The expanding versions of $\mathcal{A}$ and $\mathcal{B}$ are stable equilibrium states, and correspond to de Sitter scale factor expansion histories.  Several interesting orbits exist, which originate near an unstable ``radiation-like" point, for example $\mathcal{G}_{+}$ or $\mathcal{H}_{+}$, and evolve towards one of these accelerated expansion points. In Figure \ref{figure:compact orbits} four such orbits are presented. 
Extremely fine tuned initial conditions are required for a trajectory to pass close to the non-analytic, unstable expanding matter point $\mathcal{K}_{+}$ and evolve toward either of the de Sitter late-time attractors. It is more natural for trajectories to begin near one of the radiation like points, $\mathcal{G}_{+}$ or $\mathcal{H}_{+}$ and asymptote towards one of the de Sitter points. This feature of this class of models will be clarified in Section \ref{sectExpHist}.
\subsection{Non-compact phase-space analysis of a HS model}\label{noncompAnalysis}

Integrating the cosmological dynamical system, an expansion history for the Hu-Sawicki model is obtained, which we can compare its predictions for the Hubble parameter, deceleration parameter, total density and equation of state parameter with those of the Concordance model.  In order to calculate the expansion history, a non compact phase space must be constructed, such that the Hubble rate $H$ is no longer a dynamical variable.  We, therefore, make use of the following non-compact coordinates:


\begin{align}\label{ncvariables}
\tilde{x}&= \frac{\dot{f'}}{f'}\frac{1}{H},~~~~\tilde{v}=\frac{1}{6}\frac{R}{H^{2}}, \\ \tilde{y}&=\frac{1}{6}\frac{f}{f'}\frac{1}{H^{2}},~~~~ \tilde{\Omega}_{m} =\frac{1}{3}\frac{\rho_{m}}{f'}\frac{1}{H^{2}}.           
\end{align}
The coordinate transformation between compact and non-compact variables is as follows:
\begin{align}
\tilde{x} &= 2\frac{{x}}{{Q}}\;, \nonumber\\
& \label{comp2noncomp}\\
\tilde{u}&=\frac{{u}}{{Q}^{2}}\;,\nonumber
\end{align}
where $\tilde{u}$ represents all of $\tilde{y},\tilde{v}$ and $\tilde{\Omega}$ in turn.

The dimensionality of the dynamical system has been reduced as there is no dynamical variable which represents the normalised volume expansion. The Friedmann equation at (\ref{modfried2}) can be reshuffled to give a constraint equation in terms of the above variables:
\begin{equation}
1=\tilde{\Omega} + \tilde{v} -\tilde{x} -\tilde{y}~.
\end{equation}
Since we are interested in integrating the system with respect to redshift, the differential equations corresponding to the non-compact dynamical system can be obtained by differentiating the non-compact variables (\ref{ncvariables}) with respect to redshift, $z$:
\begin{align}
\frac{d\tilde{x}}{dz}=&\frac{1}{(z+1)}\left[  \left( -1+3\,w \right) \tilde{\Omega}+{\tilde{x}}^{2}+ \left( 1+\tilde{v} \right) \tilde{x} -2\tilde{v}+4\tilde{y}\right] , \label{Neqx}\\
& \nonumber \\
\frac{d\tilde{y}}{dz}=&-\frac{1}{(z+1)}\left[{\tilde{v}\tilde{x}{\Gamma}-\tilde{x}\tilde{y}+4\,\tilde{y}-2\,\tilde{y}\tilde{v}}\right] ,\\
& \nonumber \\
\frac{d\tilde{v}}{dz}=&-\frac{\tilde{v}}{(z+1)}\left[ { \left( \tilde{x}{\Gamma}+4-2\,\tilde{v} \right) }\right] ,\label{Neqv}\\
& \nonumber \\
\frac{d\tilde{\Omega}}{dz}=&\frac{1}{(z+1)}\left[{\tilde{\Omega}\, \left( -1+3\,w+\tilde{x}+2\,\tilde{v} \right) }\right] ,\label{NeqQ}
\end{align}
where $\Gamma$ is still given by (\ref{gamma}), due to the fact that the relationship between $v$ and $y$, and $\tilde{v}$ and $\tilde{y}$ is preserved during the coordinate transformation from compact to non-compact variables. 

In terms of the dynamical variables, we also have an expression for the dimensionless expansion rate of the universe, $h$, given by the Raychaudhuri equations,
\begin{equation}\label{numericH}
\frac{dh}{dz} = \frac{h}{z+1} \left[{ \left( 2-\tilde{v} \right) }\right]\;, 
\end{equation}
where $h=\frac{H}{H_{0}}$.

The four interesting \emph{non-boundary} points and their coordinates in the compact and non-compact phase spaces are tabulated below in Table \ref{nonboundary points}, along with their stability and their scale factor evolution, in a universe dominated by dust.  The purpose of this finite analysis is to produce an expansion history for the particular HS model under investigation with which the $\Lambda$CDM model can be compared. It is therefore only important to consider the \emph{expanding} version of the fixed points. That is to say, we need only look at the non-compact fixed points corresponding to the \emph{expanding} versions of the compact fixed points obtained earlier. 
\begin{table*}
\centering
\begin{tabular}{l p{2.4cm} p{2cm} c c}\hline \hline
Point & Non-Compact $(\tilde{x},\tilde{y},\tilde{v},\tilde{\Omega})$ & Compact $(Q,x,y,v,\Omega)$ & Stability $(w=0)$ & Scale factor solution \\ \hline
$\tilde{\mathcal{A}}_{+}$ & $[0, 1, 2, 0]$          & $[\frac{\sqrt{2}}{2}, 0, \frac{1}{2}, 1, 0]$ & \small{Attractor}&$a(t)=a_{0}e^{H_{0}(t-t_{0})}$\\
$\tilde{\mathcal{B}}_{+}$ & $[ 1, 0, 2, 0]$ & $[ \frac{2}{3}, \frac {1}{3}, 0, \frac{8}{9}, 0]$ & \small{Attractor}&$a(t)=a_{0}e^{H_{0}(t-t_{0})}$\\
$\tilde{\mathcal{G}}_{+}$      & $[-1, 0, 0, 0]$        & $[2,-1,0,0,0]$ & \small{Repellor} &$ a(t)=a_{0}\left( 2H_{0}(t-t_{0})+1 \right)^{\frac{1}{2}} $\\
$\tilde{\mathcal{H}}_{+}$ & $[ 1-3w, 0, 0, 2-3w]$ & $\left[ \frac{2}{3(w-1)},  \frac{3w-1}{3(w-1)}, 0,\right.$ $\left.0,-\frac{4}{9} \frac{3w-2}{(w-1)^{2}}\right]$ & \small{Saddle} &$ a(t)=a_{0}\left( 2H_{0}(t-t_{0})+1 \right)^{\frac{1}{2}} $  \\
$\tilde{\mathcal{K}}_{+}$ &$\left[0,\frac{1}{2},\frac{1}{2},1\right]$&$\left[\frac{\sqrt{6}}{3},0,\frac{1}{3},\frac{1}{3},\frac{2}{3}\right]$&\small{Unstable spiral}&$a(t)=a_{0}\left(\frac{3}{2}H_{0}\left(t-t_{0}\right)+1\right)^{\frac{2}{3}}$ \\ \\ \hline
\end{tabular}
\caption{There exist two stable equilibrium points for which the scale factor increases exponentially into the future; $\tilde{\mathcal{A}}_{+}$ and $\tilde{\mathcal{B}}_{+}$. It follows that these states can be associated with the late time accelerated expansion of the universe attributed to an effective $f(R)$ Dark Energy. The scale factor evolution of the points, $\tilde{\mathcal{G}}_{+}$ and $\tilde{\mathcal{H}}_{+}$ depend on the square root of time, and therefore represent ``radiation" -like universes. The non-analytic matter point is included, as it also appears in the non-compact phase space on the plane $\tilde{y}=\tilde{v}$.}
\label{nonboundary points}
\end{table*}
\section{Expansion History for the Hu-Sawicki model, with $n=1$, $c_{1}=1$}\label{sectExpHist}
As already pointed out, the utility of the finite dynamical systems analysis is to provide a fast numerically stable integration of the cosmological equations linking the early and late time evolution (represented by fixed points in the phase-space of these models).  This expansion history can then be compared to the $\Lambda$CDM model in order to determine whether it provides a good alternative to the standard model. We will see below that this depends critically on the choice of parameter values and the initial conditions.
\subsection{Initial conditions}
The expansion history for a universe governed by the HS model with $n=1$ and $c_{1}=1$ is calculated, by performing an integration of the differential equations representing the dynamical system (\ref{Neqx}) $-$ (\ref{NeqQ}). The model was parametrised based on the considerations of \cite{Hu2008} wherein fiducial restrictions were placed on the relationship between $c_{1}$ and $c_{2}$:
\begin{equation}\label{ab}
\frac{c_{1}}{c_{2}}\approx 6 \frac{\Omega_{\Lambda}}{\Omega_{m}},
\end{equation}
to control the ratio of matter density to the cosmological constant via the parameters $c_{1}$ and $c_{2}$. As is done in \cite{Hu2008}, the following values for the densities at the present epoch are used:
\begin{align}\label{HUvals1}
\Omega_{\Lambda}=0.76,~~~~~\Omega_{m0}=0.24~.
\end{align}
The initial values (at the present epoch) for the Ricci scalar, $R$, and the derivative of the function, $f'(R)\equiv f_{R}$, can be given by 
\begin{align}\label{initial R and g2}
R_{0} &\approx m^{2}\left( \frac{12}{{\Omega}_{m0}} - 9 \right), \nonumber \\
f_{R0} & \approx 1-g_{R0}=1 -n\frac{c_{1}}{c_{2}^{2}}\left( \frac{12}{{\Omega}_{m0}} - 9 \right)^{-n-1},
\end{align}
for $|g_{R0}|\ll1$. Substituting (\ref{HUvals1}) into (\ref{initial R and g2}), we have:
\begin{equation}
R_{0}=41m^{2}=41\lambda H_{0}^{2}.
\end{equation}

In order to calculate the initial values for the dynamical variables $(\tilde{x},\tilde{y},\tilde{v},\tilde{\Omega})$, expressions for each are required in terms of the quantities $q$, $r$ and $h$ (with $r$ and $h$ defined by (\ref{dimensionlessR})), for which the initial values can be obtained by using (\ref{initial R and g2}) and (\ref{HUvals1}). With $n=1$, $c_{1}=1$ we obtain:
\begin{equation}
f'(r_{0})=\frac{c_{2}r_{0} (c_{2}r_{0}+2\lambda)}{(c_{2}r_{0}+\lambda)^{2}}\;,~r_{0}=6(1-q_{0})h_{0}^2\;,
\end{equation}
from which we can obtain the deceleration parameter:
\begin{equation}
q_{0}=\frac{6h_{0}^{2}-r_{0}}{6h_{0}^{2}}\;.
\end{equation}
It follows that, in terms of the Ricci scalar, the Hubble parameter, the deceleration parameter and $c_{2}$ and $C$, the dynamical variables are:
\begin{eqnarray}
&\tilde{v}_{0}=1-q_{0}=\frac{r_{0}}{6h_{0}^{2}}\;,~\tilde{\Omega}_{0}=\frac{\Omega_{m0}(z_{0}+1)^{3}}{h_{0}^{2}f'(r_{0})}\;, \nonumber\\ & \tilde{y}_{0}=\frac{1}{6}\frac{r_{0}(r_{0}c_{2}+\lambda)}{h_{0}^{2}(r_{0}c_{2}+2\lambda)}\;,~\tilde{x}_{0}=\Omega_{0}+v_{0}-y_{0}-1\;.
\end{eqnarray}
$\Omega_{0}$ denotes the initial value of the dynamical variable $\Omega$, and is not to be confused with $\Omega_{m0}$, which represents the matter density parameter of the universe at the present epoch.
Using (\ref{ab}) and (\ref{initial R and g2}), and the fact that $m^{2}\equiv CH_{0}^{2}$, we find the following values for the parameters\footnote{$f(R)$ theories commonly suffer singularities in the expansion history, where the first derivative of the Hubble parameter diverges. The parameters chosen above enable a singularity free expansion history within the redshift range which we are interested. \cite{Diego2015} details theoretical constraints on the parameters of this model, as well as investigating certain observational constraints.}:
\begin{equation}\label{HUvals}
n=1\;,~c_{1}=1\;,~c_{2}=\frac{1}{19}\;,~\lambda=0.24
\end{equation}
and the following initial values:
\begin{equation}\label{HUinits}
h_{0}=1\;,~ r_{0} = 9.840
\end{equation}
\begin{eqnarray}
&\tilde{x}_{0,HS}=-0.339\;,~\tilde{y}_{0,HS}=1.246\;,\nonumber\\
&\tilde{v}_{0,HS}=1.640\;,~\tilde{\Omega}_{0,HS}=0.267\;.
\end{eqnarray}
The above set of coordinates correspond to the \emph{present epoch},  $\mathbf{z_{0}=0}$, as calculated from the parameter values and constraints outlined in \cite{Hu2008}.

The values for the present epoch in the compact phase space are found to be:
\begin{eqnarray}\label{compactz0}
&Q_{0,HS}= 0.719\;,~x_{0,HS}=-0.122\;,~y_{0,HS}=0.644\;,\nonumber\\
&v_{0,HS}=0.847\;,~\Omega_{0,HS}=0.138\;.
\end{eqnarray}

Below, in Figure \ref{todayBox}, an orbit in the 3D \emph{compact phase space} is presented, which has its initial values at $(Q_{0},{x}_{0},{v}_{0})$  as given by (\ref{compactz0}).
\begin{figure}[th!]
\centering
\includegraphics[width=0.57\textwidth]{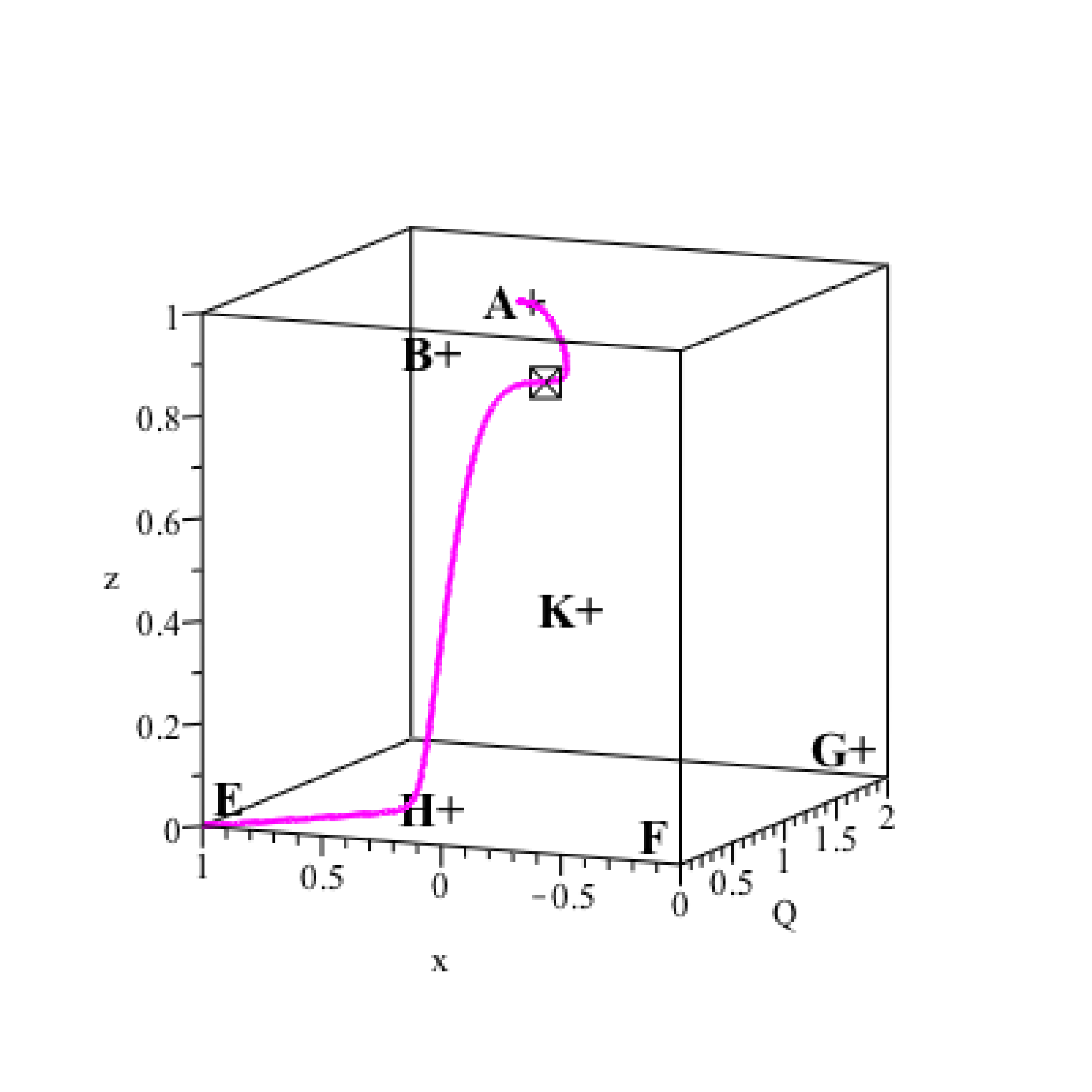} 
\caption{The fixed points represented by $\mathcal{A}_{+}$, $\mathcal{B}_{+}$, $\mathcal{G}_{+}$, $\mathcal{H}_{+}$ and $\mathcal{K}_{+}$ in the above plot correspond to the compact fixed points given in Table \ref{sols}. The matter point $\mathcal{K}_{+}$  lies on the plane $y=v$ for which $\Gamma$ is undefined. The crossed square indicates the point which corresponds to the present epoch as given by (\ref{compactz0}), \emph{today} (\textbf{$z_{0}=0$}), for the model defined by the parameters defined in \cite{Hu2008}. It can be seen that this point is close to the de Sitter stationary solution $\mathcal{A}_{+}$. From $\boxtimes$ the orbit evolves forward in time toward $\mathcal{A}_{+}$. In its past, it passes by the unstable radiation-like stationary state, $\mathcal{H}_{+}$.  Preceding the point $\mathcal{H}_{+}$, the orbit evolves from the unstable static universe phase state $\mathcal{E}$.}
\label{todayBox}
\end{figure}
\subsection{Comparing the Hu-Sawicki Model ($n=1,c_{1}=1$) with $\Lambda$CDM}
\begin{figure*}
\begin{center}
\includegraphics[width=0.3\textwidth]{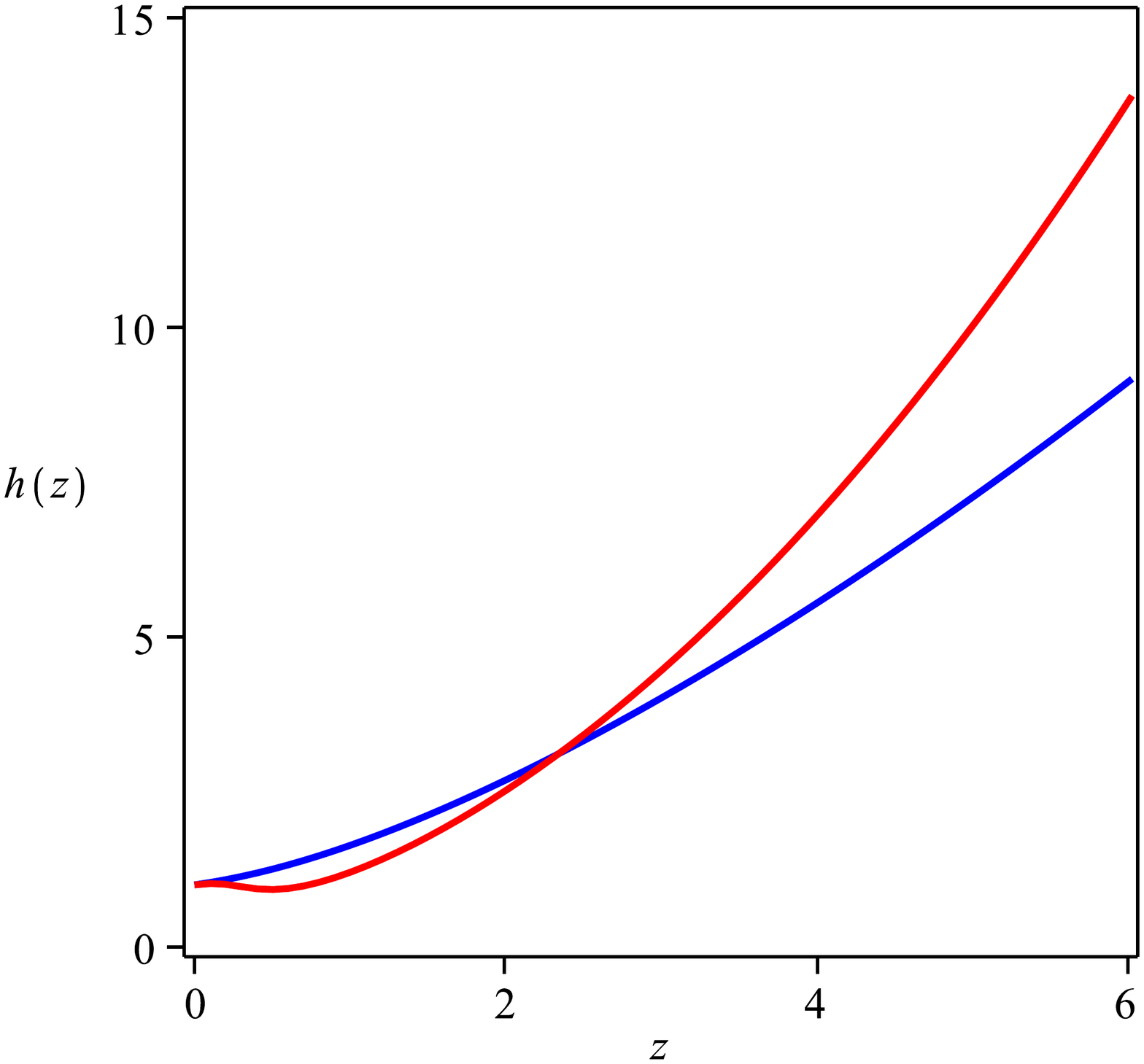}
\includegraphics[width=0.3\textwidth]{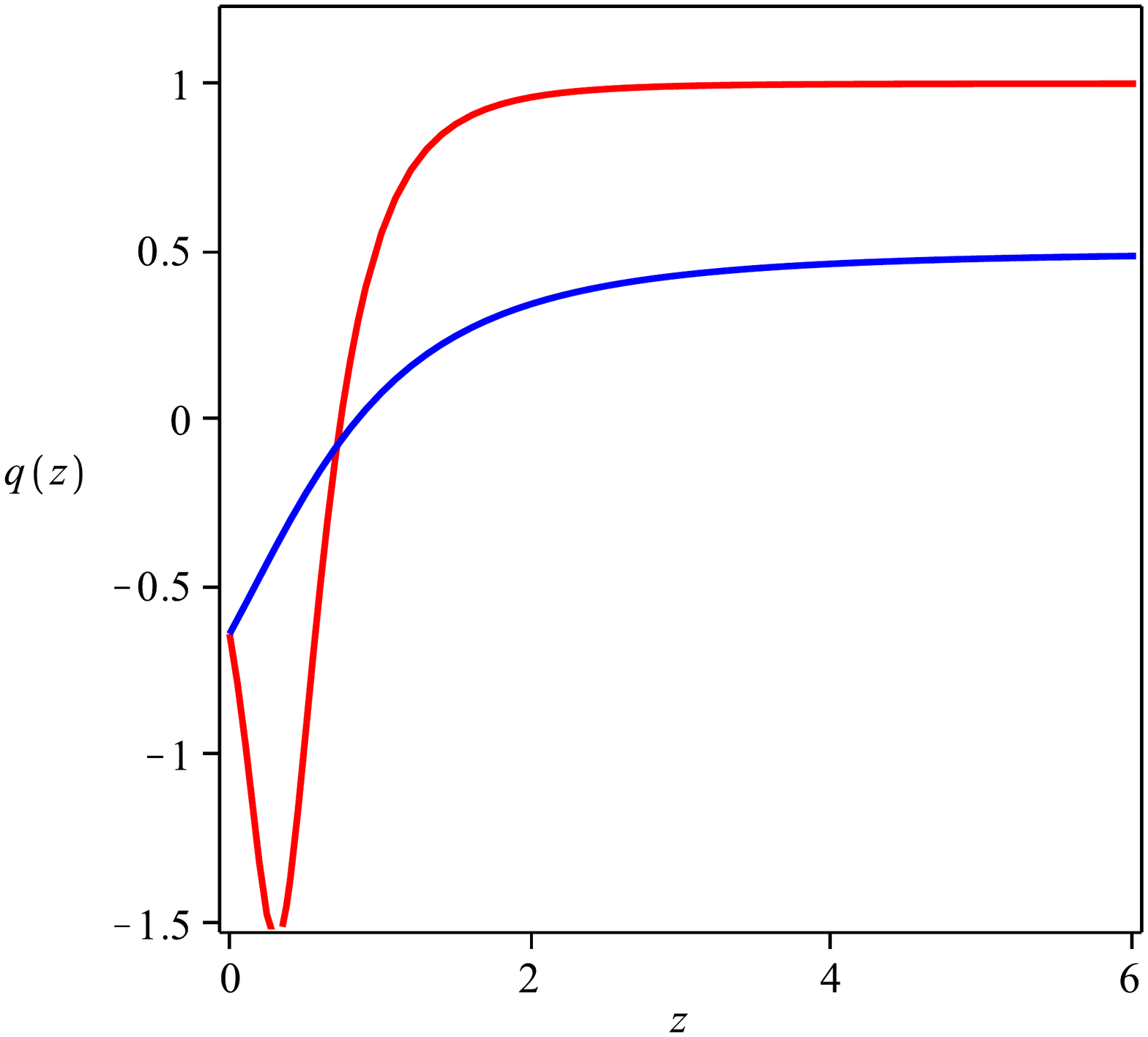}
\includegraphics[width=0.3\textwidth]{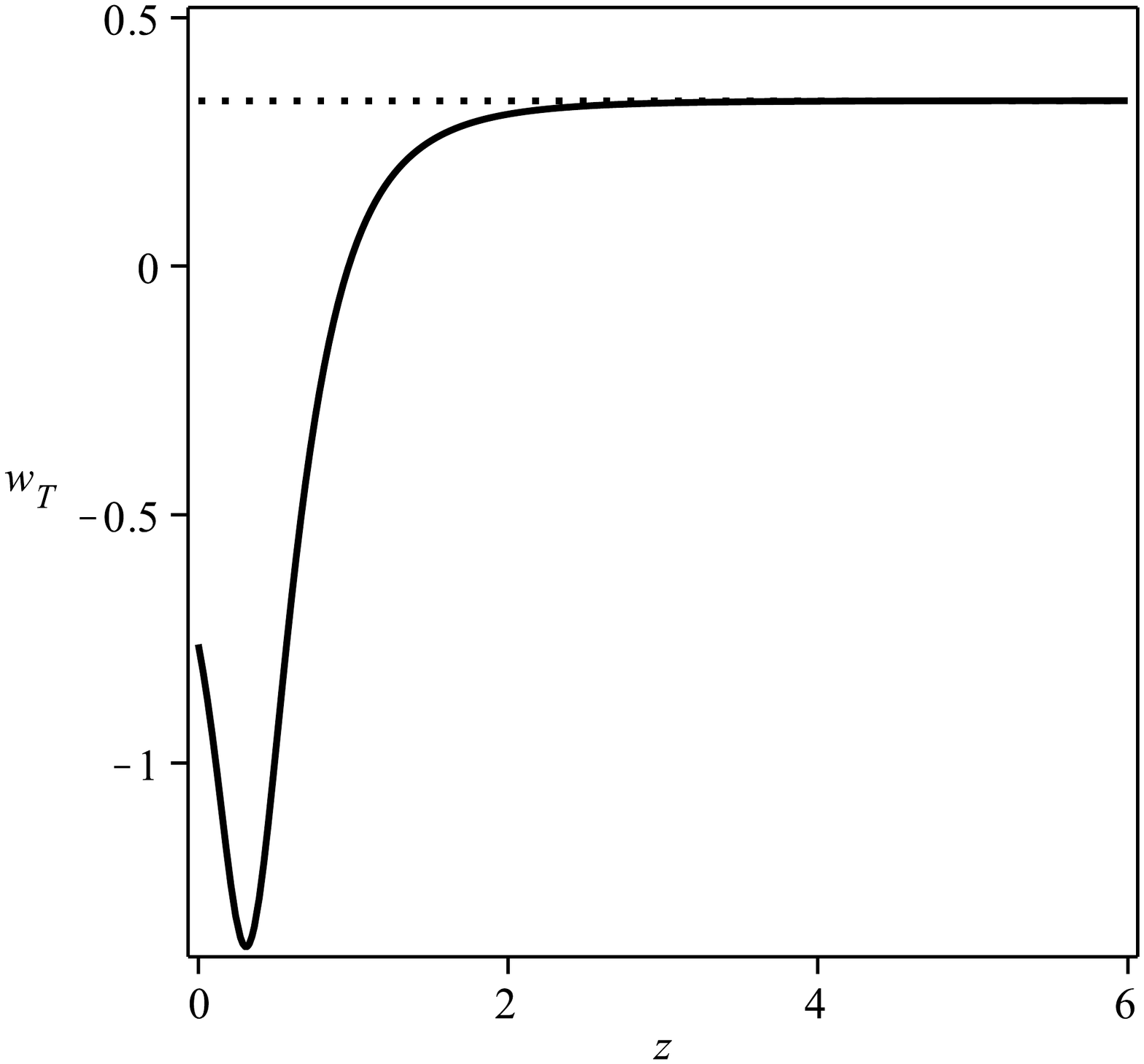}
\end{center}
\caption{Above, we plot in red the Hu-Sawicki model $(n=1,c_{1}=1,c_{2}=1/19,C=0.24)$, where initial conditions are fixed at $z=0$, and in blue, the $\Lambda$CDM model. In (a) the dimensionless Hubble parameter for $\Lambda$CDM given by equation (\ref{lcdmhubble}) is compared with the dimensionless Hubble rate obtained from the integration of the equations extracted from the non-compact dynamical systems analysis (the solution to the dynamical system (\ref{Neqx}) $-$(\ref{numericH})). At very low redshifts, the two models coincide for a short interval, but after around $z=0.05 $ they deviate substantially. 
In (b), the deceleration parameter for $\Lambda$CDM given by equation (\ref{lcdmDEC}) is compared with the deceleration parameter determined by the solution of the dynamical system at (\ref{Neqx}) $-$(\ref{numericH}), given by equation (\ref{varDEC}). The deceleration parameter determined by this specific  HS model supports late time acceleration as well as a stabilised deceleration at higher redshifts. However, there is a large discrepancy in the values for $q$ predicted by the HS model and that predicted by $\Lambda$CDM for this choice of parameters. (c) shows the redshift evolution of the total equation of state parameter determined by the HS model. It exhibits odd behaviour; at high redshifts it tends towards a constant value of $\frac{1}{3}$ indicating radiation domination-like behaviour, even though the entire analysis was performed assuming a dust filled universe of $w=0$. At low redshifts it decreases sharply to indicate the domination of a fluid with negative pressure, which reaches a minimum at -1.37 and begins to increase towards the present epoch to about -0.76. This is consistent with the dynamical systems analysis which shows an early time radiation-like repeller, no matter point, and a late time stable de Sitter phase state, as a result of the fluid with equation of state $w \approx -0.76$.
}
\label{todayHuHubble}
\end{figure*}

Figure \ref{todayHuHubble}  shows the redshift evolution of the dimensionless Hubble parameter and the deceleration parameter for the specific HS model considered above, in comparison with a $\Lambda$CDM model parametrised by the same values for $\Omega_{m}$ and $\Omega_{\Lambda}$ as above, i.e. $z_{0}=0$ or ``today".  
\subsubsection{Hubble parameter, $h$}
The dimensionless Hubble rate for the $\Lambda$CDM model is givel by:
\begin{equation}\label{lcdmhubble}
h(z) = \sqrt{\Omega_{m0}(1+z)^{3}+\Omega_{\Lambda}}~.
\end{equation}
and is compared to the solution of the differential equation for $h(z)$ given by (\ref{numericH}).
\subsubsection{The deceleration parameter, $q$}
The deceleration parameter for the $\Lambda$CDM model is given by the following expression:
\begin{equation}\label{lcdmDEC}
q = \frac{1}{H^{2}}\left(\frac{1}{2}\Omega_{m0}(1+z)^{3}-2\Omega_{\Lambda}\right),
\end{equation}
where as the deceleration parameter calculated for this specific HS model is given in terms of the dynamical variables as
\begin{equation}\label{varDEC}
q=1-\tilde{v}.
\end{equation}
\subsubsection{Total equation of state parameter $w_{T}$}
\begin{figure}[!h]
{\includegraphics[width=0.45\textwidth]{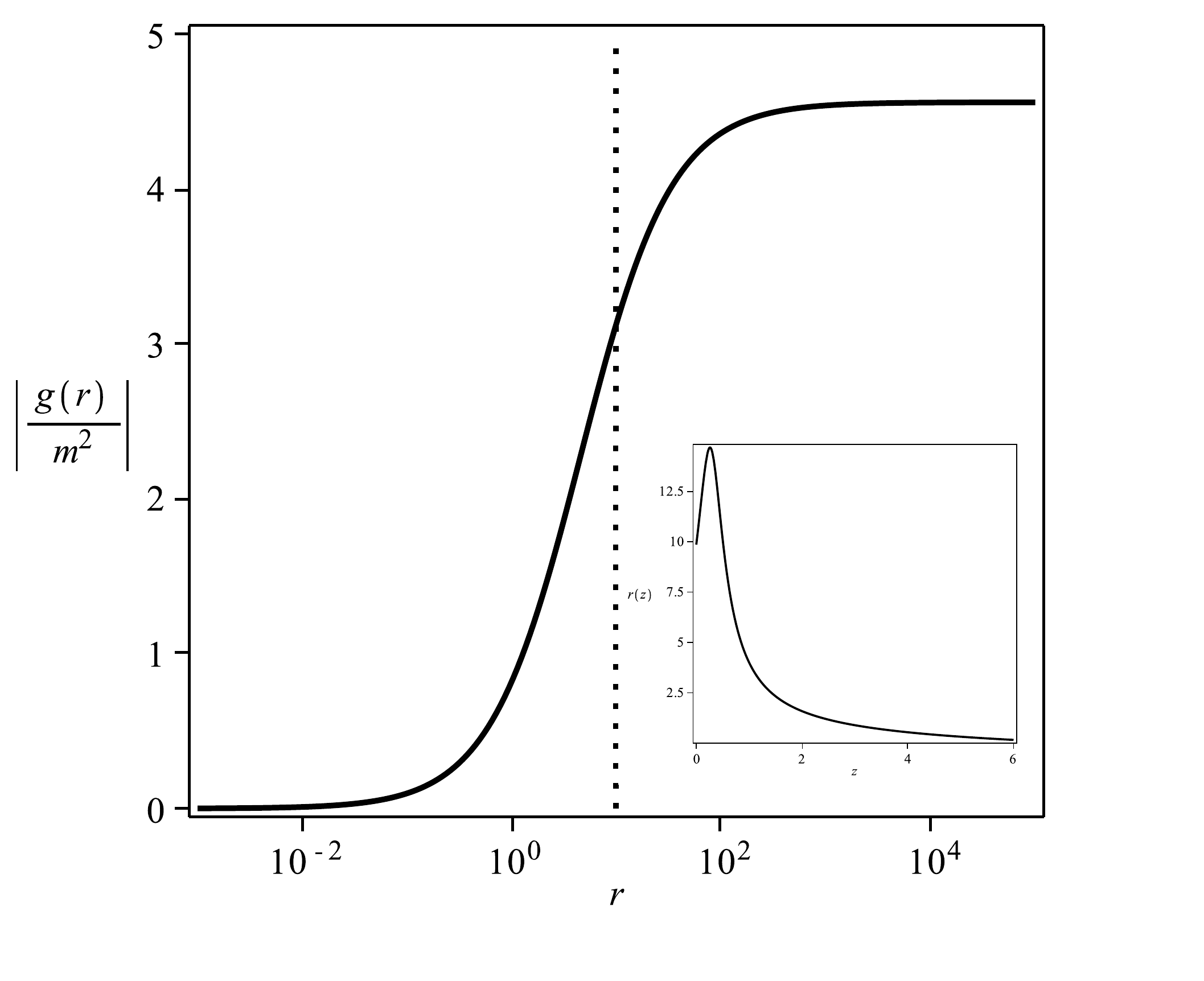}}
\caption{The main plot shows the behaviour of the correction. To fulfil its purpose, $g(R)$ is constructed to tend toward zero as $r\rightarrow 0$ in order to be compatible with GR, which is very well tested at solar system scales, and it should approach a constant value fro large $r$ in order to mimic the observed cosmological constant behaviour, which is well described, so far, by the $\Lambda$CDM model \cite{Hu2008}. The plateau of constant $g(R)$ offers a simulated cosmological constant term to the gravitational Lagrangian. The dotted black line indicates the initial value of the Ricci scalar, $R/H_{0}^{2}(z=0)$ . The inner plot shows the Ricci scalar defined as $r=R/H_{0}^{2}$ as a function of redshift.}
\label{figg_z0}
\end{figure}

It is interesting to consider the redshift evolution of the total equation of state parameter, $w_{T}$:
\begin{equation}
w_{T} = \frac{p_T}{\rho_{T}}=\frac{p_{eff}}{\rho_{m}+\rho_{eff}}.
\end{equation}
Using the trace and Friedmann equations, this can be written in terms of the dynamical systems variables defined in (\ref{ncvariables}) above:
\begin{equation}
w_{T} =\frac{1}{3}(1-2\tilde{v})
\end{equation}
The right panel in Figure \ref{todayHuHubble} shows the behaviour of the total equation of state parameter. It is interesting to note that this total equation of state parameter asymptotes toward the value of $w=\frac{1}{3}$, indicating a universe which is dominated, for most of its history, by a curvature fluid  exhibiting ``radiation-like" behaviour , and only \emph{now}, at \emph{very} low redshifts does a change in the form of the dominant energy density take place. This result is counterintuitive, owing to the fact that the entire analysis was performed with respect to a dust only universe. However, it is consistent with the dynamical systems analysis which have unstable radiation like phase states from which the corresponding solution orbit begins. This model produces a late time negative equation of state which is consistent with the dynamical systems analysis showing an approach toward a universe having scale factor which evolves exponentially with time. However, this result is not consistent with idea that the HS model approximation holds throughout the expansion history of the universe, as it predicts that a radiation fluid energy density dominates even at relatively low redshifts. This is a highly unsatisfactory match to observations and, of course, $\Lambda$CDM predictions. This result is specific to the model considered, where $n=1, c_{1}=1$. Choosing the initial parameter values to be exactly equal to their $\Lambda$CDM values today $(z_{0}=0)$ is what compromises the agreement between this model and the $\Lambda$CDM model. To obtain a better match, an appropriate adjustment of these initial values from their corresponding $\Lambda$CDM values is required. In order to understand why this is so, consider a plot of the correction $g(R)$ generated by the model with initial values specified at (\ref{HUinits}) for $\mathbf{z_{0}=0}$, i.e. the $\Lambda$CDM values corresponding to ``today", in Figure \ref{figg_z0}.

In order for the model investigated to mimic $\Lambda$CDM behaviour, that is, exhibit a GR+\emph{cosmological constant} nature, the initial value of the Ricci scalar should lie on the plateau corresponding to a constant value of $g(R)$. However, for the specific model and density parameters above, the initial value of $R/H_{0}^{2}$ at $z_{0}=0$, the present epoch, is $9.84$ as stated in (\ref{HUinits}). This value of $R/H_{0}^{2}$ is not high enough to place the correction \emph{initially} on the plateau, and therefore it does not allow the model ($n=1,c_{1}=1$) to mimic $\Lambda$CDM behaviour.  In fact, as can be seen in Figure \ref{figg_z0}, the Ricci scalar determined by this model, for initial parameter values given by (\ref{HUinits}), \emph{decreases} with redshift. This indicates that integrating from $z_{0}=0$ only drives the value of $g(R)$ further away from its $\Lambda$CDM plateau limit.
\subsection{Initial conditions at $\mathbf{z_{0}}=20$}
\begin{figure}[h!]
 {\includegraphics[width=0.45\textwidth,angle=0]{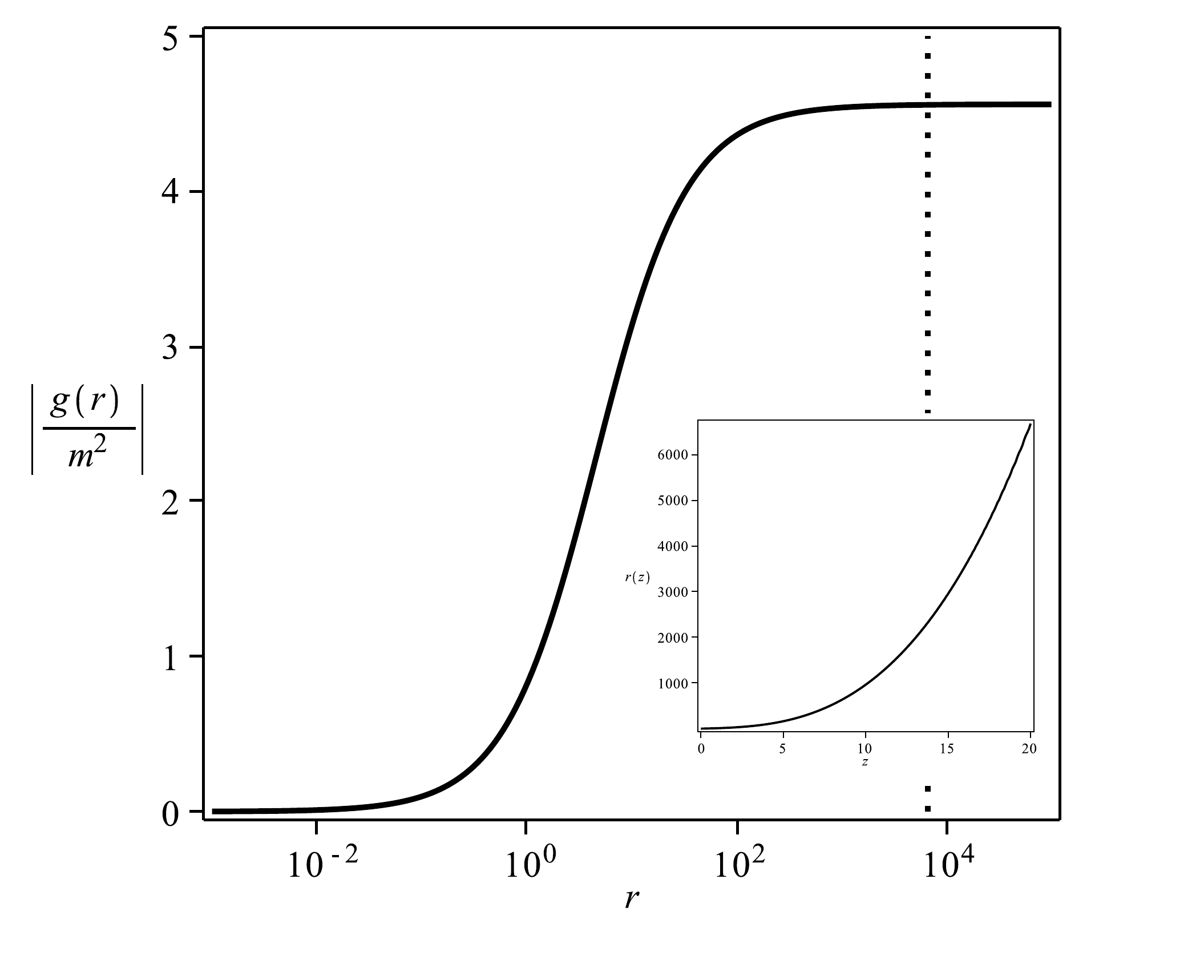}}
\caption{We show the correction, when initial conditions are fixed at $z=20$. In this case the initial value of $g(R)$ sits on the constant valued plateau of $g(R)$ which represents a simulated cosmological constant.The dotted black line indicates the initial value of the Ricci scalar, $R/H_{0}^{2}(z=20)$ . The inserted panel gives the redshift evolution of the expansion normalised Ricci scalar $r=R/H_{0}^{2}$, which is now increasing with redshift as expected, leading to a past expansion history very close to $\Lambda$CDM.}
\label{figg_z10}
\end{figure}

\begin{figure}[h!]
\centering
\includegraphics[width=0.55\textwidth]{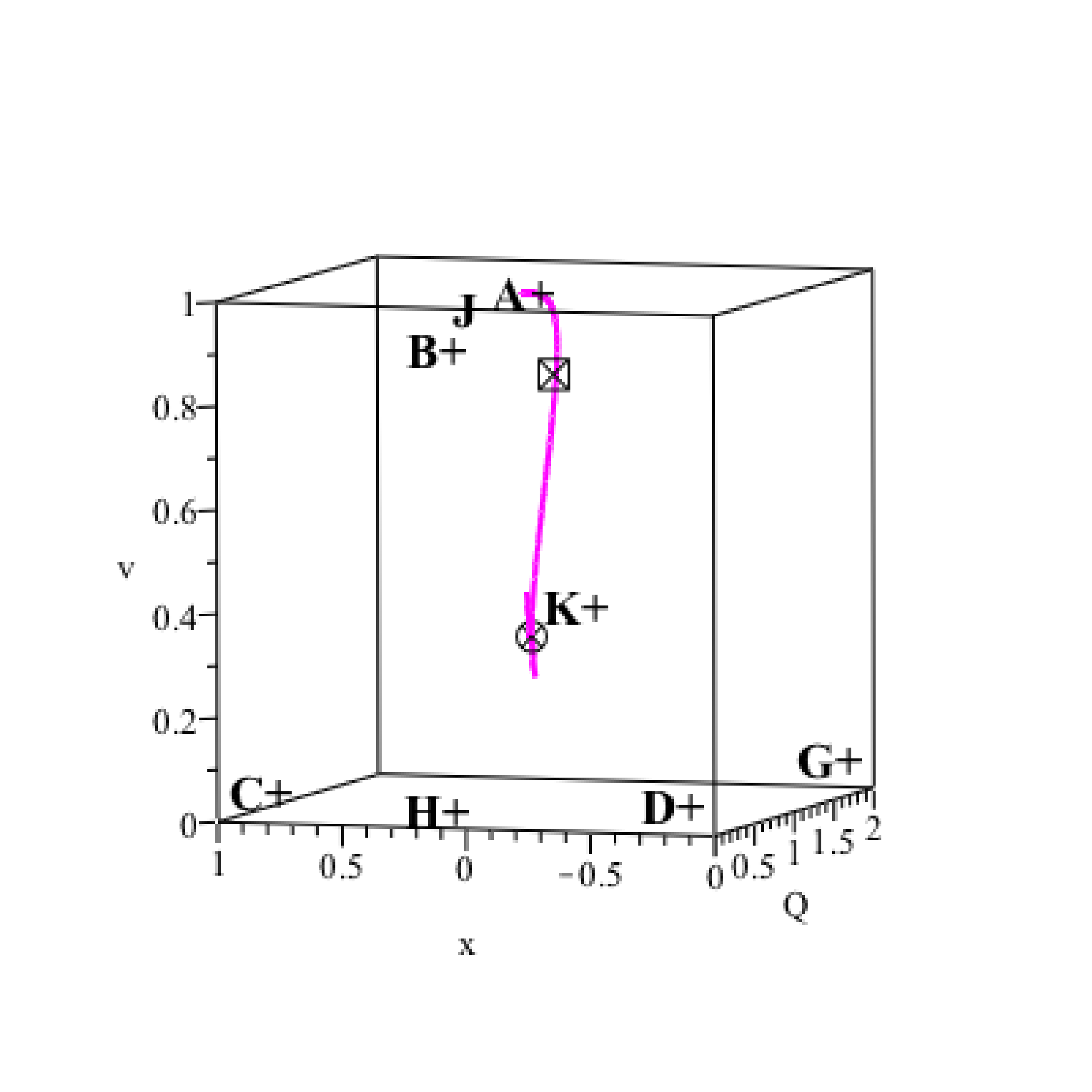} 
\caption{The above plot shows the solution trajectory of the compact dynamical systems equations, integrated from the initial values specified at (\ref{HUinitsz20}) for $z_{0}=20$. The crossed circle symbol represents the starting point ($z_{0}=20$) and the crossed square symbol represents the coordinate values at the present epoch ($z_{0}=0$). The fixed point $\mathcal{K}_{+}$ represents the matter fixed point at $(x,y,v,\Omega)=(0,\frac{1}{2},\frac{1}{2},1)$. The fixed point $\mathcal{A}_{+}$ represents a de Sitter point. This trajectory clearly begins extremely close to the matter point, {is affected by} the spiral nature of the neighbourhood of $\mathcal{K}_{+}$ on the $Q$ - $v$ plane and then evolves neatly toward the de Sitter point. This shows an expansion evolution which is similar to $\Lambda$CDM, in that it evolves from a state which is matter dominated to a state which is filled with a fluid causing exponentially accelerated late time expansion.}
\label{z20trajectory}
\end{figure}

To remedy this, we can choose initial values which increases the value of $r$ so that it sits comfortably on the plateau of $g(R)$, therefore generating an effective cosmological constant. This can be easily done by fixing our initial redshift to be $z_{0}=20$ - see Figure \ref{figg_z10}.
In this case, the initial values for the Ricci scalar, $R_{0}$ and the non-compact dynamical variables are :
\begin{eqnarray}
&h_{0}=47.153\;,~r_{0}= 6677.040\;,~\tilde{x}_{0,HS}=0.0\;,\nonumber\\
&\tilde{y}_{0,HS}=0.5\;,~\tilde{v}_{0,HS}=0.501\;,~\tilde{\Omega}_{0,HS}=0.999\;. \label{HUinitsz20}
\end{eqnarray}
The above initial values are extremely close to the matter point $\mathcal{K}_{+}$ , which is consistent with the fact that, at higher redshifts, we expect the dust fluid will dominate the equation of state of the universe. To illustrate this point, the solution trajectory with initial values given by (\ref{HUinitsz20}) is presented in Figure \ref{z20trajectory}.
\begin{figure*}
\includegraphics[width=0.32\textwidth]{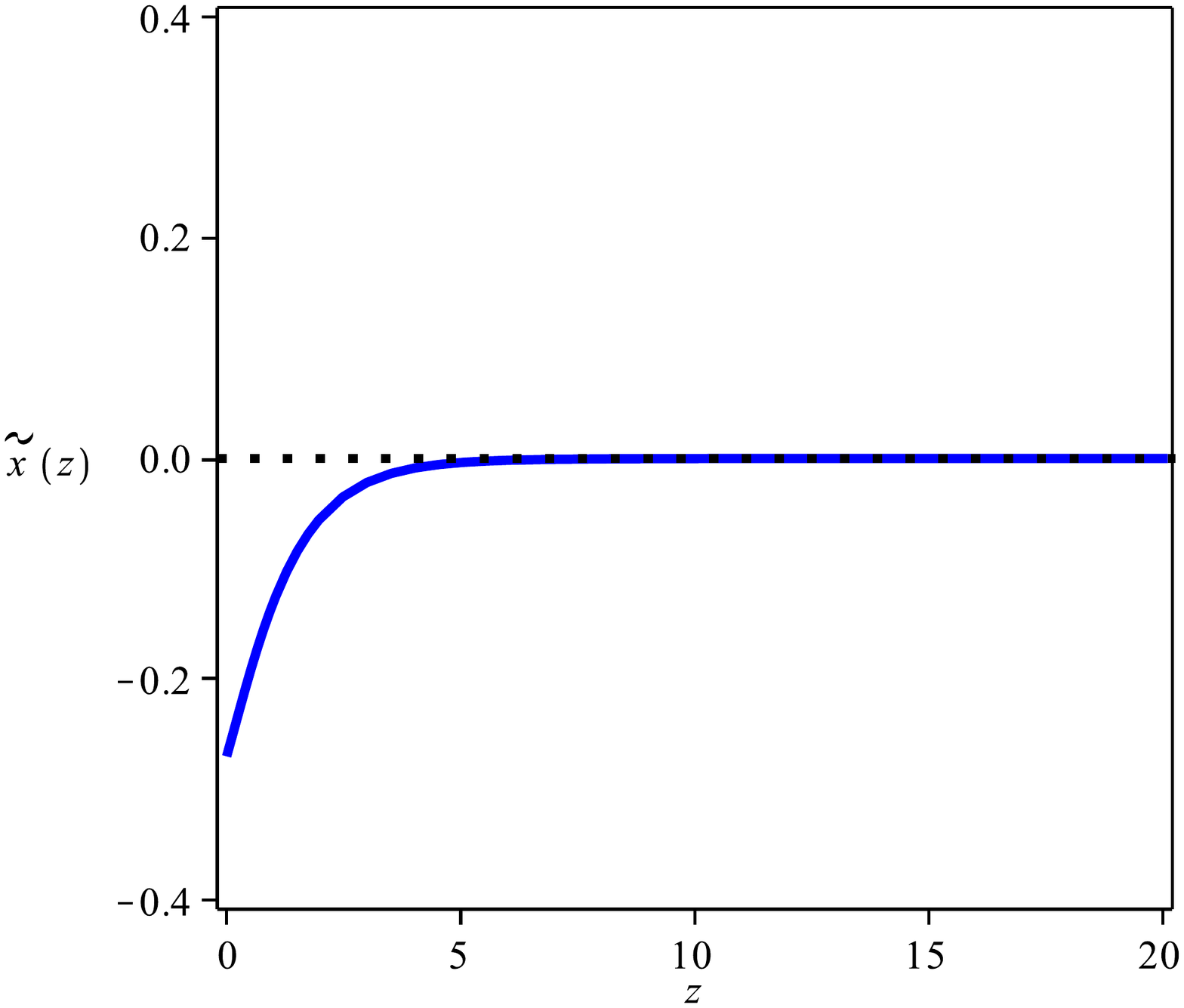}
\includegraphics[width=0.32\textwidth]{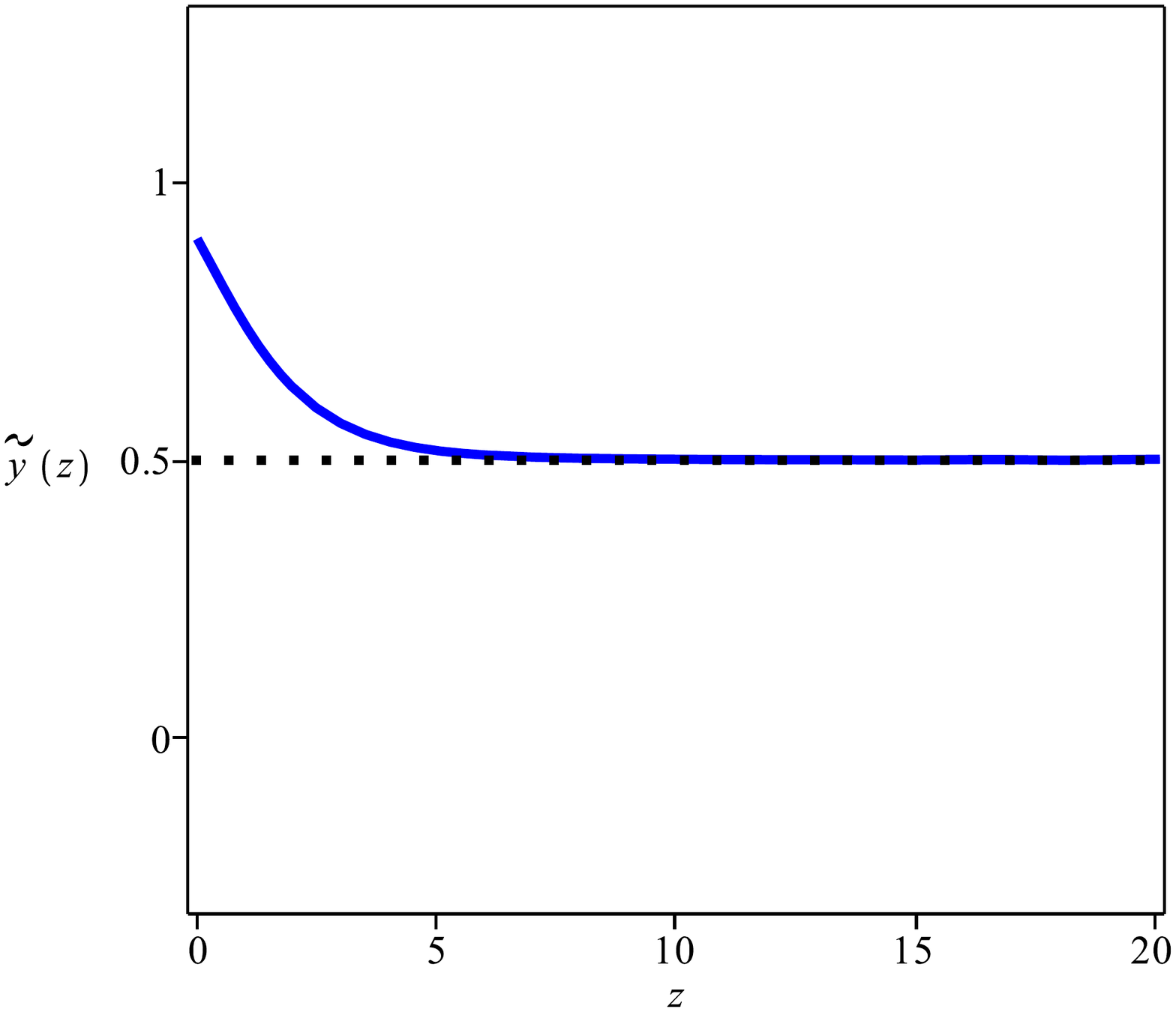}
\includegraphics[width=0.32\textwidth]{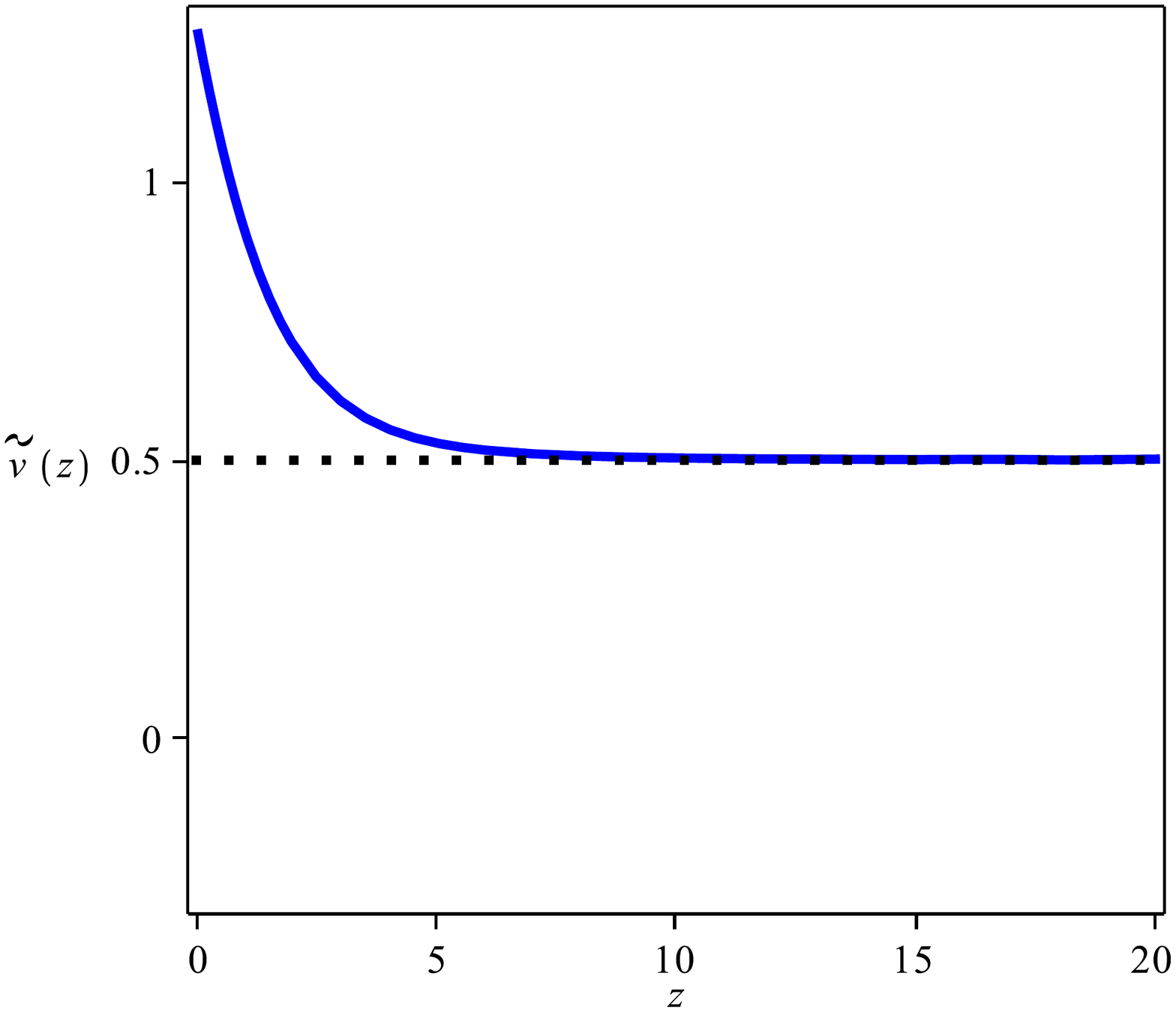}  
\caption{The $(\tilde{x},\tilde{y},\tilde{v})$ coordinates of the matter points, $\mathcal{K}_{+}$ are $\left(0,\frac{1}{2},\frac{1}{2}\right)$, indicated on the plots by black dotted line.  The difference between these values and the initial values of the dynamical variables $\tilde{x},\tilde{y}$ and $\tilde{v}$ is too small to resolve at a redshift greater than 20. We therefore assert that beginning the integration at this redshift is sufficient to determine the behaviour of this model in the $\Lambda$CDM regime.}
\label{redshift evolution}
\end{figure*}

At this point, it is worth noting that the value of $z_{0}=20$ is sufficiently high in redshift to begin the integration of the dynamical system. As we increase redshift, the values of the coordinates tend closer and closer to the matter point $\mathcal{K}_{+}$, more and more quickly, to such an extent that it makes little difference whether we begin at $z=1000$ or $z=20$, as we have chosen. Figure \ref{redshift evolution} demonstrates this fact, showing the dynamical variables as a function of redshift. The asymptote of the variables toward their respective matter point, $\mathcal{K}_{+}$, coordinates is obvious.

\begin{figure*}
{\includegraphics[width=0.32\textwidth,angle=0]{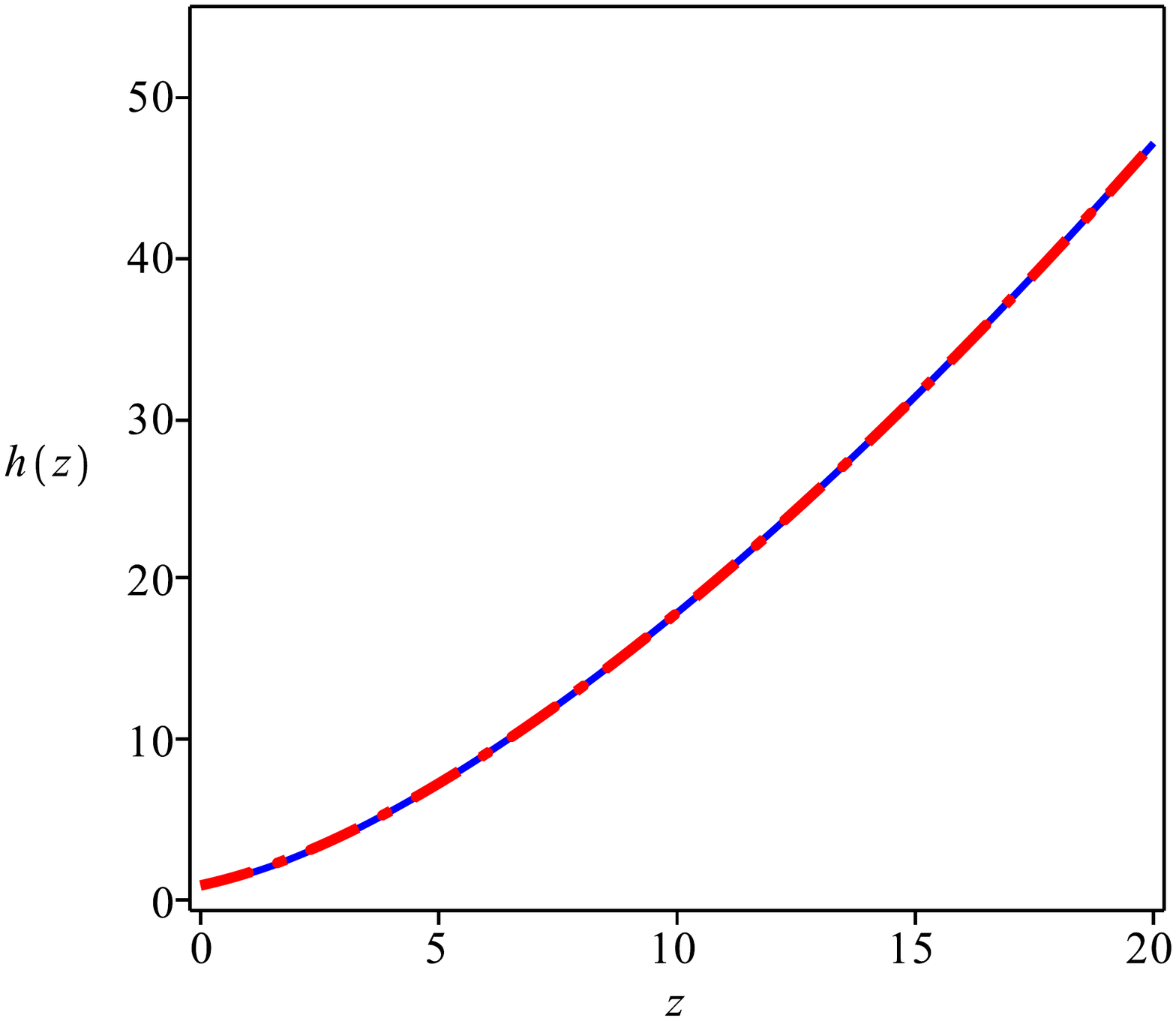}}
{\includegraphics[width=0.32\textwidth,angle=0]{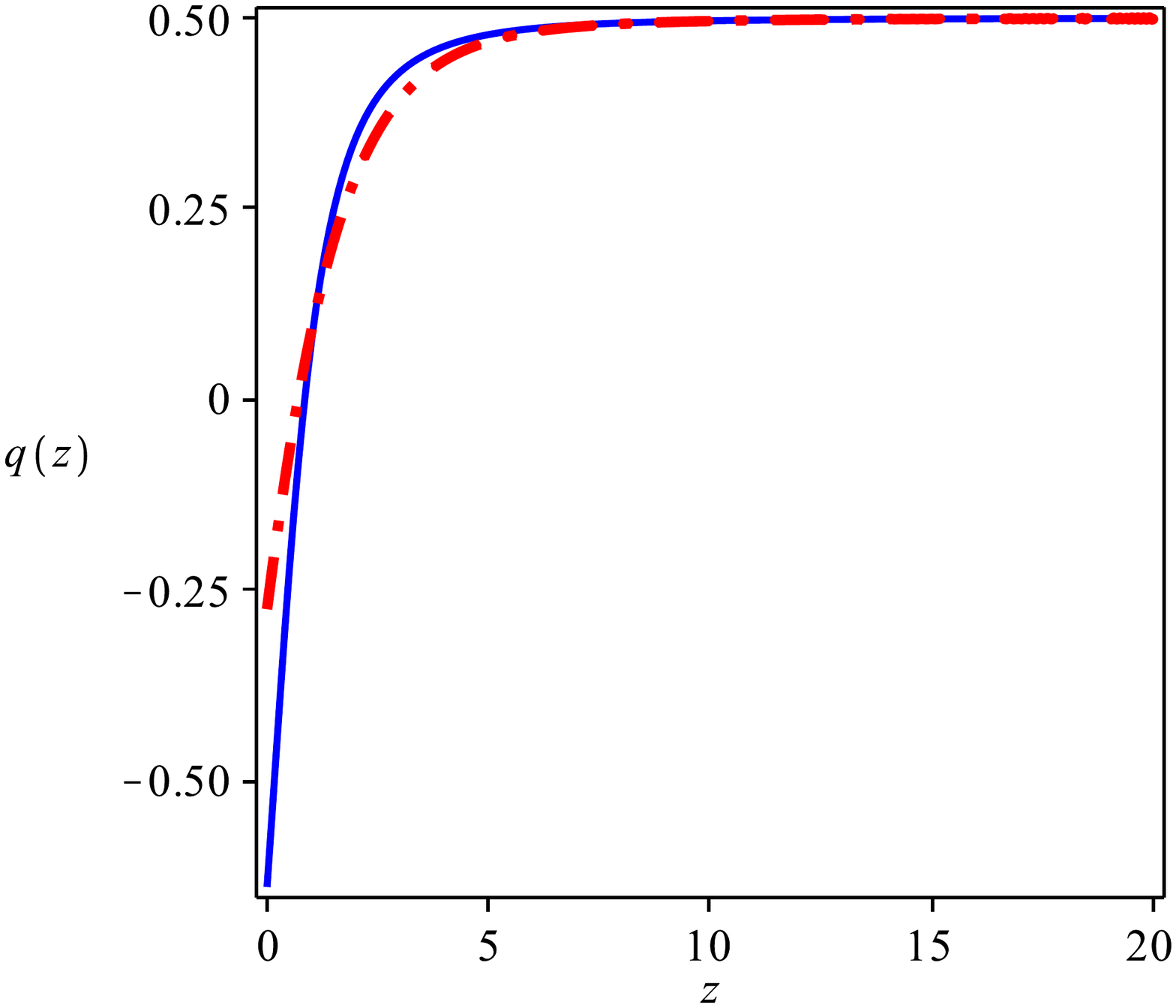}}
{\includegraphics[width=0.32\textwidth,angle=0]{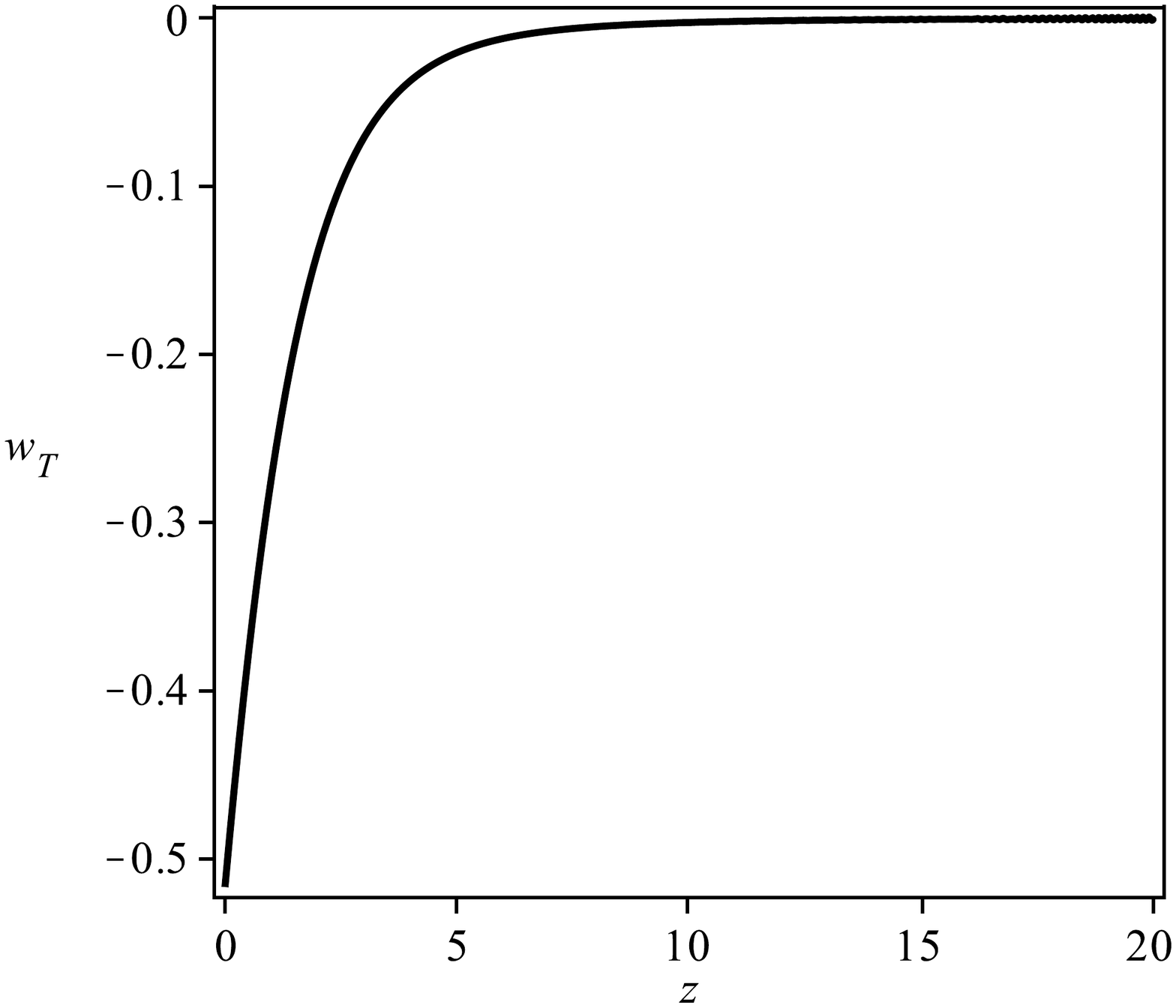}}
\caption{Here we provide a comparison of the Hu-Sawicki model with $n=1,c_{1}=1,c_{2}=1/19,C=0.24$ (red dashed line) and the $\Lambda CDM$ model (\ref{lcdmhubble}) (solid blue line), where initial conditions are fixed at $z=20$. The left panel gives the dimensionless Hubble parameter which is almost indistinguishable from $\Lambda CDM$ . The middle panel gives the deceleration parameters, giving both late time acceleration today and $q_{HS}\Rightarrow q_{\Lambda CDM} \approx \frac{1}{2}$ at high redshifts. The right panel gives a total effective equation of state which clearly asymptotes towards $w_{Total}=0$ at high redshifts (as expected) as well as tending toward a negative value in the low redshift regime, thus behaving like dark energy. These results represent a dramatic improvement on what was obtained for initial conditions at $z=0$, shown in figure 4, which was unable to produce a matter dominated evolution at high redshift.}
\label{fig_z20}
\end{figure*}

To demonstrate the effects of placing the initial value of the correction initially on its function's plateau, the corresponding plots of the dimensionless Hubble parameter, the deceleration parameter and the equation of state parameter is presented in Figure \ref{fig_z20}.

It is clear that placing the correction $g(R)$ \emph{initially} on the constant valued plateau permits the specific HS model analysed to mimic the $\Lambda$CDM model relatively well. By doing this, we have corrected the asymptote of the equation of state from $\frac{1}{3}$ to $0$, we have managed to simulate the Hubble parameter, $h$, of the $\Lambda$CDM model very closely, and there is an improvement in the early time behaviour of the deceleration parameter. From these results it can be concluded that the HS model can produce a very good simulation of the $\Lambda$CDM model, however, this occurs at the expense of the present $\Lambda$CDM values of $q$, $h$ and $r$. {In order to obtain the desired behaviour; for the cosmology to be dominated by a dust fluid in the past, the values of  $q_{\Lambda CDM}$, $h_{\Lambda CDM}$ and $r_{\Lambda CDM}$ corresponding to the present epoch $ (z=0)$ can not be used as starting values for the integration. These parameters must be adjusted to enable $g(R)$ to initially assume its plateau value.}
\subsubsection{Effects of increasing $n$}
\begin{figure}[h!]
 {\includegraphics[width=0.45\textwidth,angle=0]{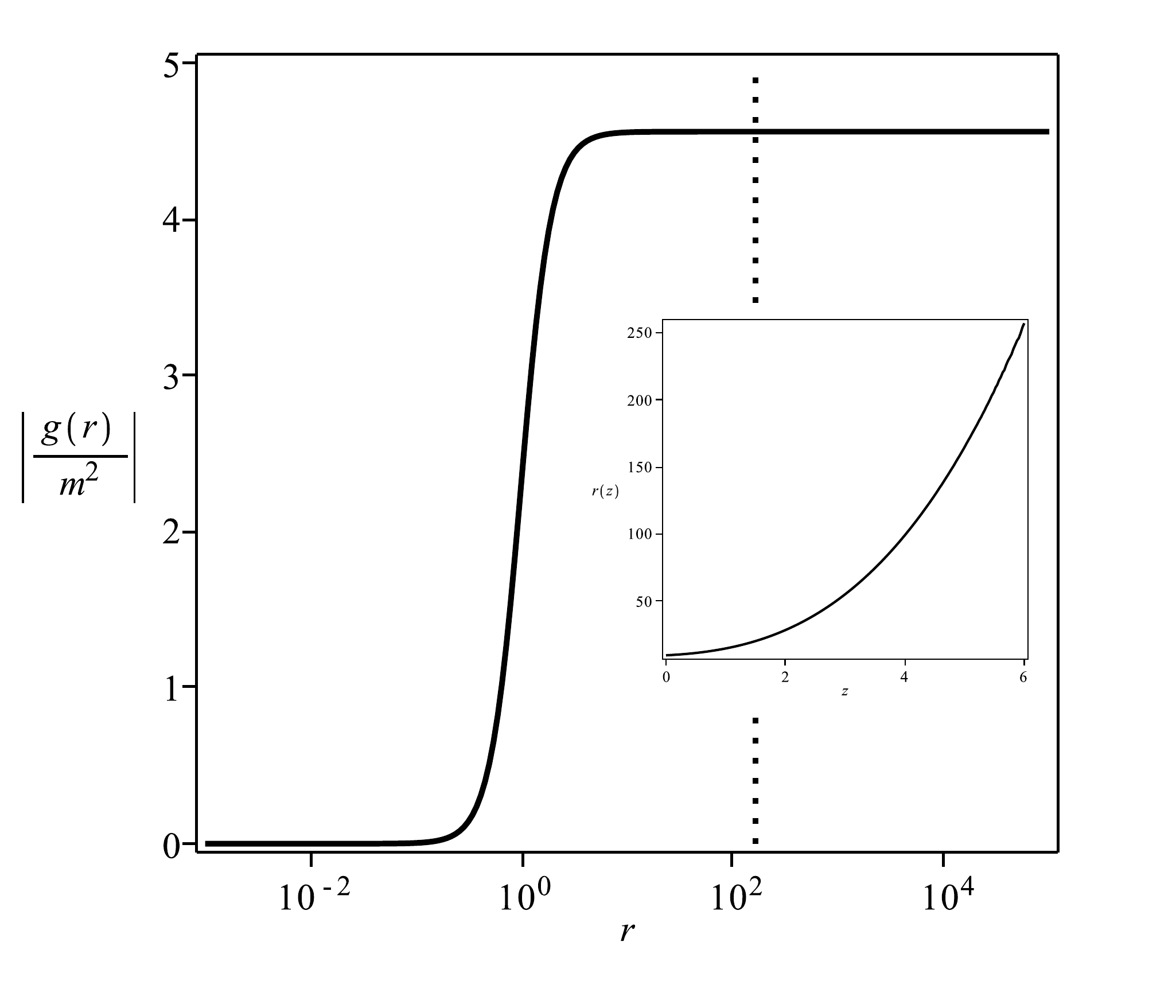}}
\caption{Above, is a plot of the correction $g(R)$ for $n=3$. It shows a significantly steeper slope, enabling a more rapid transition from the GR limit to the $\Lambda$CDM limit of the function. This slope enables the HS model (for $n>1$ ) to mimic the $\Lambda$CDM model even with initial values set at the present epoch, $\mathbf{z_{0}=0}$. However, $n \neq 1$ is not permitted by a dynamical systems analysis performed with the variables defined by (\ref{variables}) or (\ref{ncvariables}), therefore other means must be used to compute its resulting expansion history. The inserted panel shows the behaviour of the corresponding Ricci scalar.}
\label{figg_z0n3}
\end{figure}

Note that the parameter $n$ plays an important role in the derivative of $g(R)$. Large values of $n$ result in a \emph{steeper}, slope of the transition between the limiting values of $g(R)$. Therefore increasing $n$ $-$ making the transition from $g(R)\rightarrow 0$ to $g(R)\rightarrow Constant$ more rapid $-$ could place the initial value for $g(R)$ at the present epoch ($z_{0}=0$) on the constant plateau of the correction function.  Figure \ref{figg_z0n3} shows a plot of the correction, $g(R)$ for $n=3$. In this plot, it is clear that the initial value of $r$ at $z_{0}=0$ results in a value of $g(R)$ which sits well on the constant part of the function, enabling a model which closely resembles the $\Lambda$CDM model at high curvature.  

\section{Conclusions}\label{sect-conc}
In this paper we present for the first time a detailed analysis of the cosmological dynamics of the Hu-Sawicki (HS) model. In order to achieve this, we first expressed the fourth-order cosmological equations in terms of a set of generalised  dimensionless expansion normalised dynamical variables which, span the phase space of this problem.  While these equations are completely general for any $f(R)$ theory of gravity, the choice of theory is fixed by expressing the term  $\Gamma$ (which appears in the equations) in terms of the set of dynamical variables, thus closing the system of equations. In the case of the HS model, this can only be done for the particular choice of model parameters: $n=1$ and $c_{1}$. While this may appear somewhat restrictive, it is easy to show using other numerical experiments, that this choice does not lead to a loss of generality. Indeed, choosing $n > 1$, simply leads to an improved fit to the standard  $\Lambda$CDM expansion history \cite{Santos2012}. 

A compact dynamical systems analysis was performed using a positive definite normalisation, which brings any equilibrium points at infinity to the boundary of the region defined by the range of the dynamical systems variables.  From this analysis, twelve equilibrium points were identified, each having an expanding ($Q>0$) and contracting $(Q<0)$ version. Four de Sitter like stationary states were found $\mathcal{A}_{\pm}$ and $\mathcal{B}_{\pm}$, two within the expanding part of the phase space, with the other two corresponding to the collapsing versions. Focusing on the expanding versions of these points,  both $\mathcal{A}_{+}$ and $\mathcal{B}_{+}$ are \emph{stable} stationary phase states. Two unstable ``radiation like" states - $\mathcal{G}_{+}$ and $\mathcal{H}_{+}$ exist in the expanding sector of the phase space. There are several orbits which connect these ``radiation" points to the de Sitter phase states, offering trajectories which resemble the chronological evolution of the scale factor of our universe. 

On the surface $y=v$, a very interesting, non-analytic, \emph{matter-like} point, $\mathcal{K}_{+}$, was identified  which after careful analysis, turns out to be an unstable spiral point. There does exist a trajectory which evolves from $\mathcal{K}_{+}$ toward a de Sitter phase state at $\mathcal{A}_{+}$, the existence of which indicates that the Hu-Sawicki model provides a modification to gravity able to produce expansion histories consistent with the $\Lambda$CDM model.

In order to consider the expansion history generated by the HS model, a finite dynamical systems analysis was performed to confirm that the non-boundary points obtained in the compact analysis were, in fact, finite points. The finite points obtained in this analysis are the analogies of the de Sitter and ``radiation like" points, $\mathcal{A}_{\pm},\mathcal{B}_{\pm}, \mathcal{G}_{\pm}$ and $\mathcal{H}_{\pm}.$ 

It was found that for $n=1$, the expansion history generated from the initial values and parameter values specified in \cite{Hu2008} for the \emph{present epoch} resulted in a universe which is dominated until very low redshifts by an equation of state parameter equal to $\frac{1}{3}$, implying that fixing the initial conditions to be exactly those of the $\Lambda$CDM conditions today leads to an expansion history very different from $\Lambda$CDM and high redshifts ($z>1$). This is due to the fact that for $n=1$, the slope of the correction $g(R)$ defined by the HS model does not allow for the initial value of $g(R)$ to sit on the constant valued plateau of the function. It is this plateau which mimics the late time cosmological constant behaviour of the model. We showed that by setting initial conditions to be the same as  $\Lambda$CDM model at high redshift, leads to a value of $g(R)$ which is equal to its ``plateau" value and consequently leads to an expansion history which is much closer to the $\Lambda$CDM model. Moreover it produces a dimensionless Hubble parameter that is nearly indistinguishable from $\Lambda$CDM, with deviations only appearing in the dynamics of the deceleration parameter.  Most importantly, it gives an equation of state parameter which behaves like CDM for most of the expansion history, and begins to tend toward -1 to indicate the domination of an effective cosmological constant term, at low redshifts.
 
Unfortunately for values of $n$ different from $1$, the dynamical systems formalism presented here does not lead to a set of autonomous equations which can be analysed in the same way. However, we expect that the key features found in the $n=1$ case to remain the same. Never the less finding a set of variables which is able to facilitate the dynamical systems analysis a general HS model in order to perform a complete analysis of the general phase space of these models remains an interesting problem which has been partially addressed in a recent paper \cite{Sante-HS}.
\section*{Acknowledgments}
%
We would like to thank David Bacon for a comprehensive reading of the manuscript and for his useful comments.
S.K. is grateful to the NRF and the Faculty of Science, University of Cape Town for financial support.
P. K. S. D. thanks the National Research Foundation (NRF) for financial support.


\end{document}